# Oscillatory double-diffusive convection in a horizontal cavity with Soret and Dufour effects

Jin Wang[1, 2], Mo Yang[1], Ya-Ling He[3], Yuwen Zhang[2, 3] [*]

[1] *School of Energy and Power Engineering, University of Shanghai for Science and Technology, Shanghai 200093, China*

[2] *Department of Mechanical and Aerospace Engineering, University of Missouri, Columbia, MO 65211, USA*

[3] *Key Laboratory of Thermo-Fluid Science and Engineering of MOE, School of Energy and Power Engineering, Xi'an Jiaotong University, Xi'an,* Shaanxi*, 710049, China*

## ABSTRACT

Oscillatory double-diffusive convection in horizontal cavity with Soret and Dufour effects is investigated numerically based on SIMPLE algorithm with QUICK scheme in non-uniform staggered grid system. The results show that double-diffusive convection develops from steady-state convection-dominated, periodic oscillatory, quasi-periodic oscillatory to chaotic flow, and finally return to periodic oscillation as buoyancy ratio increases. Moreover, fundamental frequency and fluctuation amplitude increase with buoyancy ratio. As Rayleigh number increases, transition trendy of oscillatory convection is similar to that of buoyancy ratio. But the return of periodic oscillation from chaos is not obtained as Rayleigh number increases. As aspect ratio decreases, the oscillatory convection evolves from periodic into steady-state. In addition, fundamental frequency increases at first and then decreases while fluctuation amplitude decreases with aspect ratio.

**Keywords:** Oscillatory double-diffusive convection; Soret and Dufour effects; Chaos; Heat and mass transfer.

**Introduction**

Since its first appearance as an oceanographical topic [1], double-diffusive convection where heat and solute of different diffusivities affect simultaneously the density and fluid motion has matured into a subject with wide applications in a large variety of fields such as astrophysics [2], manufacturing [3-5] and ventilations [6,7]. Compared with nature convection only driven by thermal buoyancy, double-diffusive convection has marked differences on heat and mass transfer. As thermal and solutal buoyancies play important roles on fluid flow and heat transfer during the double-diffusive convection process, it is necessary to develop effective models and methods to better understand the double-diffusive convective mechanism. During the past several decades, many numerical and experimental studies focused on double-diffusive convection have been carried out [8-13].

As a comprehensive flow and heat transfer problem, double-diffusive convection in a typical configuration is a strong and complex nonlinear problem [10, 13]. Any difference between thermal and solutal buoyancies or diffusivities may induce convective instabilities even if the initial and boundary conditions are gravitatively stable. For several decades, researchers devoted to stationary and oscillatory nonlinear characteristics of double-diffusive convection with different thermal Rayleigh numbers, Lewis numbers, buoyancy ratios, and aspect ratios. Huppert [14] analyzed transition from conduction state to oscillatory motion followed by transition to a more complicated oscillatory motion with increasing thermal Rayleigh number for double-diffusive convection between two infinite planes. Further transition between oscillatory and steady convection was reported by Costa et al. [15] using a second order nonlinear model for two-dimensional double-diffusive convection. Khadiri et al. [16] investigated double-diffusive convection in a square porous cavity heated and salted from below based on the study of natural convection heating from below [17] and monocellular, bicellular and tricellular flows were presented and discussed in detail. Ghorayeb et al. [18] studied numerically the onset of oscillatory double-diffusive convection in a square cavity and the influence of Lewis number on the transition from steady convective flow to oscillatory flow was carried out. Chen et al. [19] expanded on Bergeon's [20] work in which only stationary onset of instability for double-diffusive convection in inclined cavity was considered and investigated oscillatory convection using linear stability analysis; the results showed two Hopf bifurcation points were obtained as aspect ratio increased. Nonlinear bifurcation analyses of double-diffusive convection in vertical enclosures were also considered by Xin et al. [21] and Bardan et al. [22].

---

[*] Corresponding author. Email: zhangyu@missouri.edu.

During the nonlinear double-diffusive convection process driven by thermal and solutal buoyancies, most studies only considered contributions of Fourier conduction and Fickian diffusion to heat and mass transfer. However, some investigations show that the Soret and Dufour effects also play significant roles during the nonlinear process [23-25], especially when the thermal and solutal buoyancies are large [26]. More studies about double-diffusive convection have been carried out recently by taking the Soret and Dufour effects into account in actual engineering transport systems. Malashetty and Gaikwad [27] investigated the Soret and Dufour effects on double-diffusive convection in an unbounded vertically stratified system using normal mode analysis. Latter, an analytical study of linear and nonlinear double-diffusive convection in coupled stress fluid considering Soret and Dufour effects was presented by Gaikwad et al. [28]. Yu et al. [29] developed lattice Boltzmann model for double-diffusive convection with Soret and Dofour effects in a horizontal shallow cavity. Nithyadevi and Yang [30] analyzed numerically double-diffusive convection of water in a partially heated enclosure with Soret and Dufour effects and the effects of heating location on heat and mass transfer were investigated. Double-diffusive convection with Soret and Dufour effects in cavity with horizontal temperature and concentration gradients was studied by Geng et al. using experimental [31] and numerical [32] methods, respectively.

For double-diffusive convection in horizontal cavity which is applied widely in many industrial processes, the flow structure and heat transfer also demonstrate complex nonlinear characteristics [14, 15, 33] as thermal Rayleigh number increases. As a nonlinear flow problem, the existing studies about double-diffusive convection in horizontal cavity are mostly concerned much more with some of the parameters or conditions such as thermal Rayleigh number [14, 15], than all of the parameters (buoyancy ratio, Prandtl number and aspect ratio). Meanwhile, Soret and Dufour effects make the nonlinear convection in horizontal cavity more complex and intricate. To predict more accurately double-diffusive convection in horizontal cavity, Wang et al. [34] developed a finite-difference numerical model of higher precision for the nonlinear convection considering Soret and Dufour effects, and systematically investigated effects of all parameters on flow pattern and heat transfer of the steady double-diffusive convection using the model [34, 35]; the results demonstrated that oscillatory convection was observed under some conditions. In the present study, numerical simulations of the oscillatory double-diffusive convection in horizontal cavity under different conditions are presented. Effects of thermal Rayleigh number, buoyancy ratio, Soret and Dufour effects, and aspect ratio on oscillatory double-diffusive convection in horizontal cavity are discussed in detail.

## 2. Problem statement and modeling

The physical model for double-diffusive convection in horizontal cavity under consideration is shown schematically in Figure 1. The horizontal cavity with aspect ratio of $A = H/W$, where H and W are height and width of the cavity respectively, is subject to vertical temperature and concentration gradients between the upper and lower horizontal walls while the two vertical side-walls are adiabatic and impermeable. In other words, the lower wall is maintained at the high temperature $T_h$ and rich concentration $c_h$; the upper wall is subject to a uniform lower temperature and concentration of $T_l$ and $c_l$. In the horizontal cavity, it is filled with a binary medium of air and solute that have an initial temperature $T_0$ and concentration $c_0$.

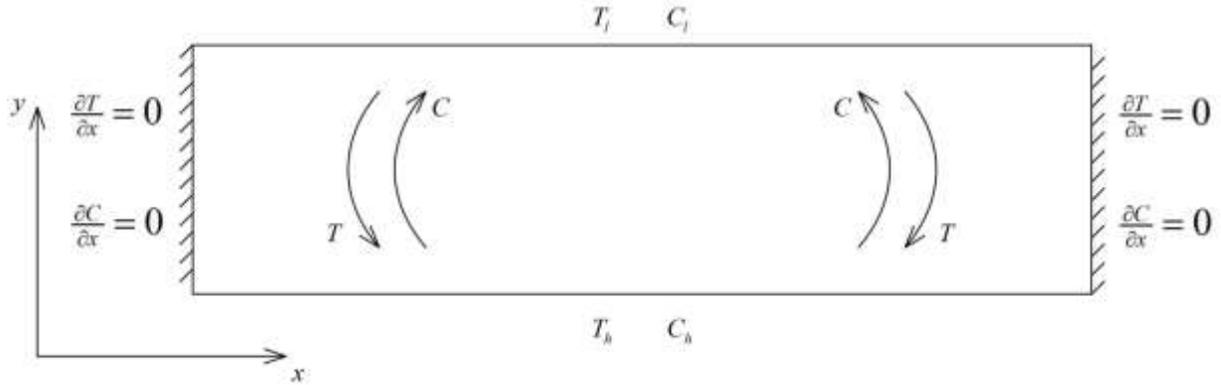

**Figure 1** Physical model for double-diffusive convection in horizontal cavity

For the two-dimensional double-diffusive convection in horizontal cavity, there is no heat generation and chemical reactions, and the effects of thermal radiation on the convection is so small that can be ignored. Meanwhile, it is assumed that all thermophysical properties of the binary medium are constants except for the density in the buoyancy term accords with the Boussinesq approximation and can be given by

$$\rho = \rho_0 \left[ 1 - \beta_T (T - T_r) - \beta_c (c - c_r) \right] \tag{1}$$

where $\rho_0$ is the medium density at the reference temperature $T_r = \dfrac{T_h + T_l}{2}$, and concentration $c_r = \dfrac{c_h + c_l}{2}$. And $\beta_T$, $\beta_c$ are the thermal and solute volumetric expansion coefficients, respectively.

The governing equations of mass, momentum, energy and species for laminar double-diffusive convection in horizontal cavity with Soret and Dufour effects can be written as following dimensionless forms, measuring lengths and time in terms of $(X, Y) = \dfrac{(x, y)}{W}$ and $\tau = \dfrac{\alpha t}{W^2}$ [34].

$$\frac{\partial U}{\partial X}+\frac{\partial V}{\partial Y}=0 \tag{2}$$

$$\frac{\partial U}{\partial \tau}+U\frac{\partial U}{\partial X}+V\frac{\partial U}{\partial Y}=-\frac{\partial P}{\partial X}+\Pr(\frac{\partial^2 U}{\partial X^2}+\frac{\partial^2 U}{\partial Y^2}) \tag{3}$$

$$\frac{\partial V}{\partial \tau}+U\frac{\partial V}{\partial X}+V\frac{\partial V}{\partial Y}=-\frac{\partial P}{\partial Y}+\Pr(\frac{\partial^2 V}{\partial X^2}+\frac{\partial^2 V}{\partial Y^2})+Ra\cdot\Pr\left[\theta-\frac{1}{2}+N_C(C-\frac{1}{2})\right] \tag{4}$$

$$\frac{\partial \theta}{\partial \tau}+U\frac{\partial \theta}{\partial X}+V\frac{\partial \theta}{\partial Y}=(\frac{\partial^2 \theta}{\partial X^2}+\frac{\partial^2 \theta}{\partial Y^2})+D_f(\frac{\partial^2 C}{\partial X^2}+\frac{\partial^2 C}{\partial Y^2}) \tag{5}$$

$$\frac{\partial C}{\partial \tau}+U\frac{\partial C}{\partial X}+V\frac{\partial C}{\partial Y}=\frac{1}{Le}\left[(\frac{\partial^2 C}{\partial X^2}+\frac{\partial^2 C}{\partial Y^2})+S_r(\frac{\partial^2 \theta}{\partial X^2}+\frac{\partial^2 \theta}{\partial Y^2})\right] \tag{6}$$

where

$$(U,V)=\frac{(u,v)W}{\alpha}, \quad \theta=\frac{T-T_l}{T_h-T_l}, \quad C=\frac{c-c_l}{c_h-c_l}, \quad P=\frac{W^2(p+\rho_0 gy)}{\rho_0 \alpha^2}, \quad \Pr=\frac{\nu}{\alpha}, \quad Le=\frac{\alpha}{D},$$

$$Ra=\frac{gW^3\beta_T(T_h-T_l)}{\nu\alpha}, \quad N_C=\frac{\beta_C(c_h-c_l)}{\beta_T(T_h-T_l)}, \quad D_f=\frac{\kappa_{TC}(c_h-c_l)}{\alpha(T_h-T_l)}, \quad S_r=\frac{\kappa_{CT}(T_h-T_l)}{D(c_h-c_l)}.$$

Here the dimensionless parameters are Prandtl number $\Pr$, Rayleigh number $Ra$, Buoyancy ratio $N_C$, Dufour number $D_f$, Lewis number $Le$ and Soret number $S_r$, respectively. $\kappa_{TC}$, $\kappa_{CT}$ are the Soret and Dufour coefficients, respectively.

The initial temperature and concentration are the same as those at the upper wall. The non-slip boundary conditions are imposed over the walls, so the dimensionless boundary and initial conditions can be expressed as:

$$X=0, \; U=V=0, \; \frac{\partial \theta}{\partial X}=0, \; \frac{\partial C}{\partial X}=0 \tag{7}$$

$$X=1, \; U=V=0, \; \frac{\partial \theta}{\partial X}=0, \; \frac{\partial C}{\partial X}=0 \tag{8}$$

$$Y=0, \; U=V=0, \; \theta=1, \; C=1 \tag{9}$$

$$Y=A, \; U=V=0, \; \theta=0, \; C=0 \tag{10}$$

$$\tau=0, \; U=V=0, \; \theta=0, \; C=0 \tag{11}$$

To characterize heat and mass transfer of the double-diffusive convection in horizontal cavity driven by thermal and solutal buoyancies combining with the Soret and Dufour effects, the local Nusselt number $Nu$ and Sherwood number $Sh$ considering the Soret and Dufour effects are defined by,

$$Nu(X) = \left.\frac{\partial \theta}{\partial Y}\right|_{Y=0} + D_f \left.\frac{\partial C}{\partial Y}\right|_{Y=0} \quad (12)$$

$$Sh(X) = \left.\frac{\partial C}{\partial Y}\right|_{Y=0} + S_r \left.\frac{\partial \theta}{\partial Y}\right|_{Y=0} \quad (13)$$

where the first and second terms of Eq. (12) indicate the Fourier's conductive flux and the diffusion thermo flux due to Dufour effect, respectively. The first and second terms in Eq. (13) denote the Fick's diffusion mass flux and the thermodiffusion mass flux induced by Soret effect, respectively. In addition, the average Nusselt and Sherwood numbers can be obtained from:

$$\overline{Nu} = \int_0^1 Nu(X)dX \quad (14)$$

$$\overline{Sh} = \int_0^1 Sh(X)dX \quad (15)$$

## 3. Numerical solution and nonlinear method

The dimensionless governing differential equations for the double-diffusive convection are discretized on a staggered grid system based on the finite volume method [36]. The QUICK scheme is employed for the convective term while the transient term is discretized using Euler backward second-order implicit scheme. The SIMPLE algorithm is implemented to solve the discretized equations by iterations [36]. During the numerical iteration process, the correction of velocity field is established by the pressure correction method based on the mass continuity to obtain the real velocity field. Meanwhile, the under-relaxation method is employed for the convergence of the simulation results and the criterion on time step is established as follow,

$$\sum_{i,j} \left| \phi_{i,j}^n - \phi_{i,j}^{n-1} \right| < 10^{-6} \quad (16)$$

where $\phi$ is the generic variable that can be $U$, $V$, $\theta$, or $C$, and the superscript $n$ indicates the iteration number on time step. The subscript sequence $(i,j)$ represents the grid node.

Based on the grid and time step independence test, the non-uniform grid of $150 \times 80$ and time step of 0.002 are chosen for present numerical study. However, a denser non-uniform grid of $180 \times 100$ is employed to ensure the grid independent solution under condition of a higher $Ra$. To improve convergence under large $Ra$ conditions, the block-correction technique [37, 38] is applied to solve the momentum equation. Meanwhile, phase space and Fourier frequency spectrum methods of nonlinear analysis are also implemented to investigate the oscillatory double-diffusive convection. For phase-space trajectory of velocity, the velocity at the center of the cavity is chosen to study the nature of oscillatory behaviors from micro perspective.

## 4. Results and discussions

The validity studies of the simulation code for the double-diffusive convection with Soret and Dufour effects have been carried out by Wang et al. [34, 35]. In present study, additional test cases are also examined with $A=1$, $Le=1$ and $\Pr=0.71$ corresponding to Béghein et al. [39]. Comparisons of temperature flied, flow structure, Nusselt and Sherwood numbers have been carried out. Table 1 presents the comparison of the results obtained by the numerical code and Béghein et al. [39]. It can be seen that the agreement between the numerical and Béghein [39] solutions is very well under all conditions with a maximum relative deviation within 2%.

Table 1 Comparison of Nusselt number between Béghein and numerical solutions for double-diffusive convection

| $N_C$ | Béghein [39] | Present | Deviation (%) |
|---|---|---|---|
| 0.1 | 16.44 | 16.24 | 1.22 |
| 0.5 | 13.96 | 13.83 | 0.93 |
| 0.8 | 10.81 | 10.75 | 0.56 |
| 1.5 | 13.96 | 13.83 | 0.93 |

Numerical simulations are then performed for the double-diffusive convection of air in horizontal cavity with Soret and Dufour effects. The Prandtl and Lewis numbers are taken to be 0.71 which corresponding to air and 1.641 corresponding to volatile organic compounds (for example Propanol) [32], respectively. As oppose steady double-diffusive convection that the flow field develops from conduction-dominated to steady convection-dominated, investigations focused on oscillatory double-diffusive convection are presented in present study. It is found that the steady convection-dominated flow becomes unstable via a Hopf bifurcation and the periodic oscillatory convection flow is observed as buoyancy ratio $N_C$ or Rayleigh number $Ra$ increases [34]. It is difficult to estimate the exact critical value of $N_C$ or $Ra$ for underlying physical problems by numerical calculations because there are many factors to limit this task. However, the transition process can be investigated as the flow structure changes.

Figure 2 shows that multiple solutions of steady-state convection corresponding to monocellular counter-clockwise flow (Fig. 2a), monocellular clockwise flow (Fig. 2b), bicellular ascending flow (Fig. 2c) and bicellular descending flow (Fig. 2d) are obtained in terms of flow field, isotherms and isosolutes when $N_c$ is equal to 0.5. More investigations show the flow structure depends on the initial condition of the double-diffusive convection. In the present study, it focuses on the transition route of double-diffusive convection, so the initial condition of Eq. (11) is chosen for the following computed cases and the effects of thermal Rayleigh number, buoyancy ratio, Soret and Dufour effects, and aspect ratio on the transition route are discussed.

Figure 3 show oscillatory route of temperature distributions for $N_C = 2.0$. It can be seen that the temperature distribution with a larger stagnant zone does not remain steady-state anymore but presents self-sustained oscillation with time. It is observed clearly that the contours near the vertical walls become thinner and then denser, and then the oscillatory behaviors repeat consistently. Meanwhile, to better visualize oscillatory behaviors, phase-space trajectory of velocity, time-evolution of heat and mass transfer, phase-space trajectory and Fourier frequency spectrum of Nusselt number analysis are presented. It can be seen from Fig. 4 that the double-diffusive convection is periodic self-sustained oscillation with fundamental frequencies are 2.2498. Further investigations for different buoyancy ratios ($N_C \leq 2.0$) are carried out and similar transition routes are obtained. And it is found that flow pattern evolves from steady-state convection-dominated into periodic self-sustained oscillatory convection as $N_c$ increases.

**U**

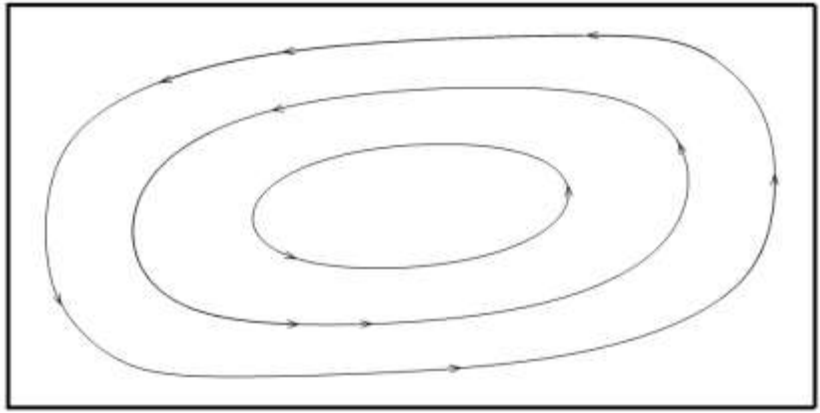

*θ*

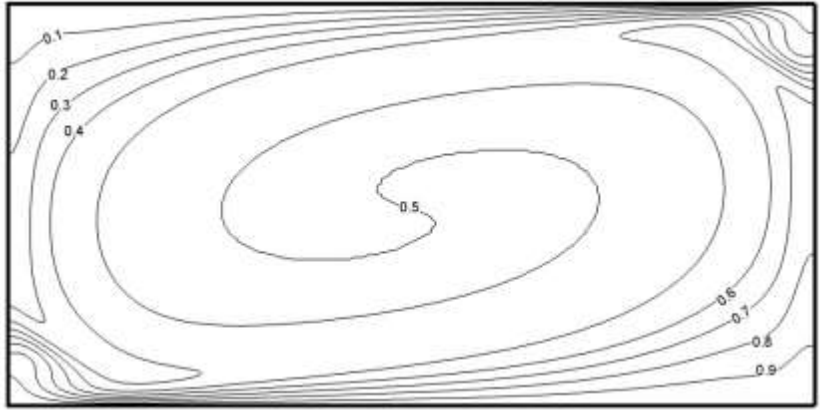

*C*

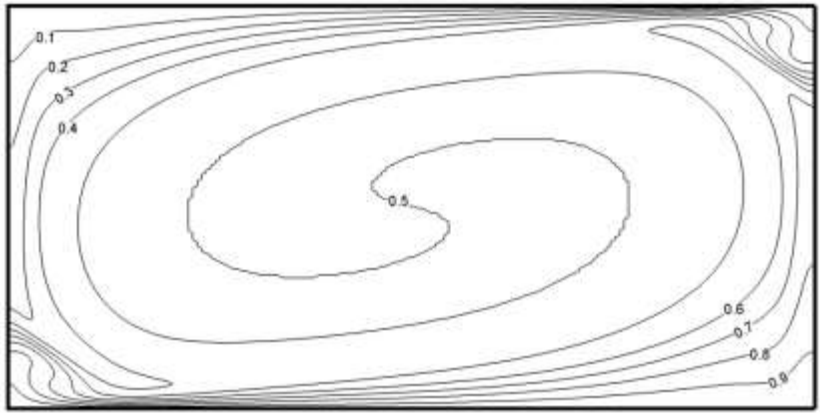

a) monocellular counter-clockwise flow

U 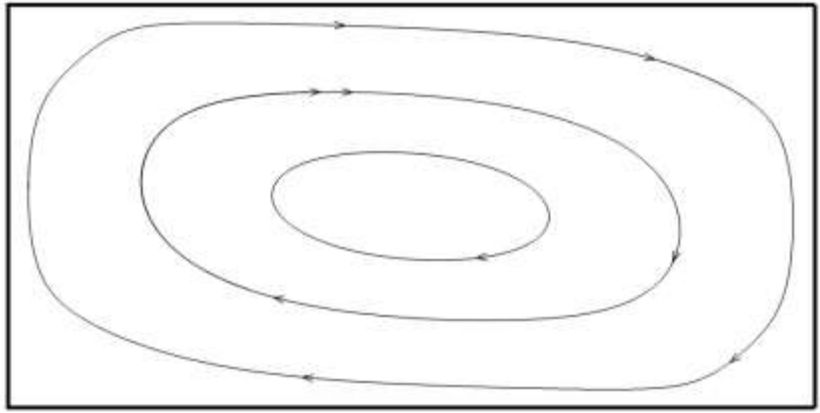

θ 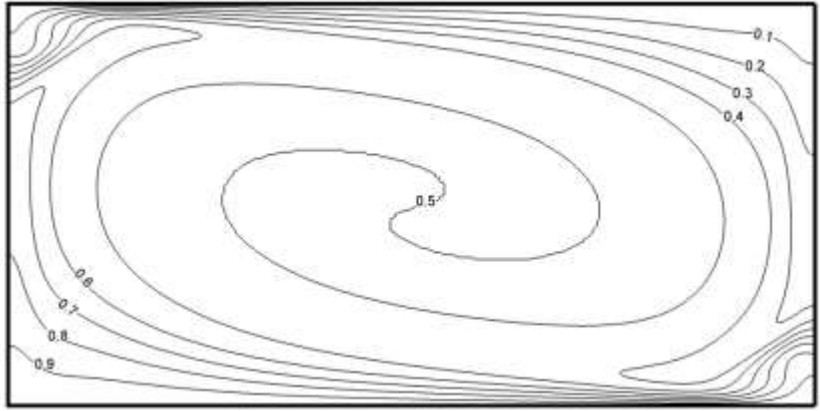

C 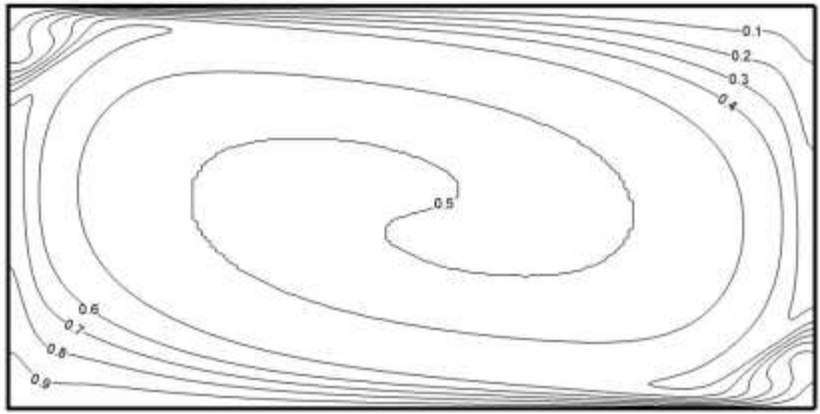

b) monocellular clockwise flow

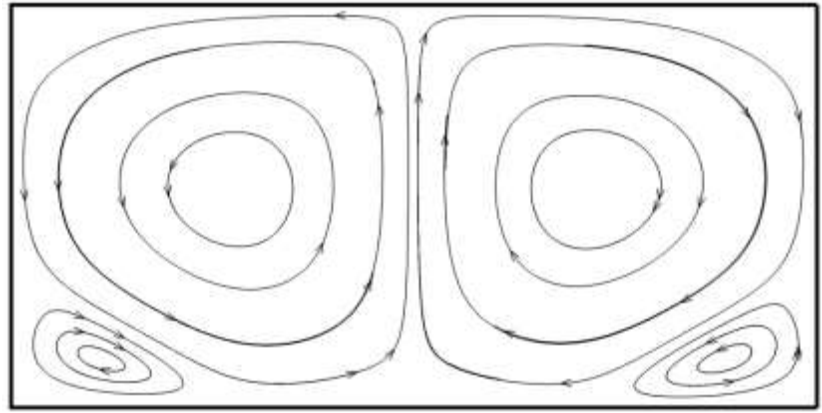

**U**

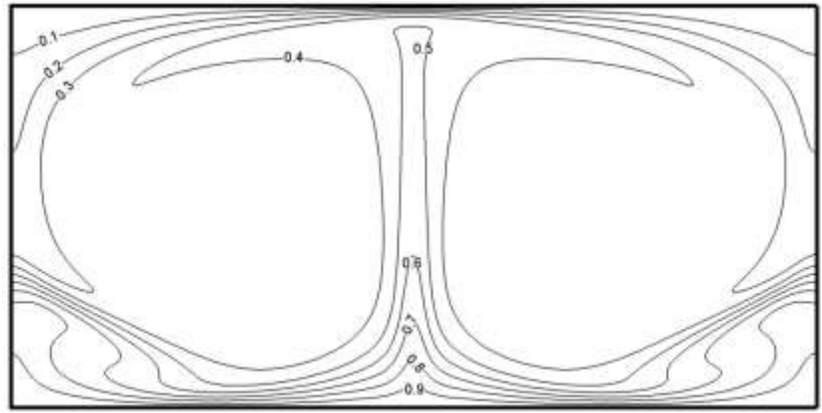

*θ*

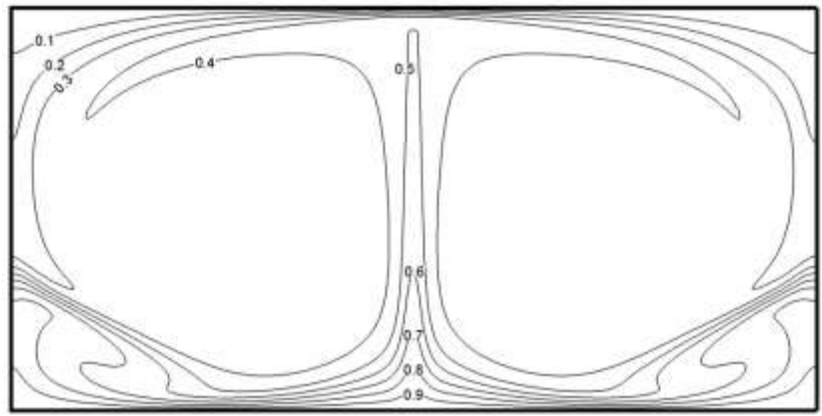

*C*

c) bicellular ascending flow

U

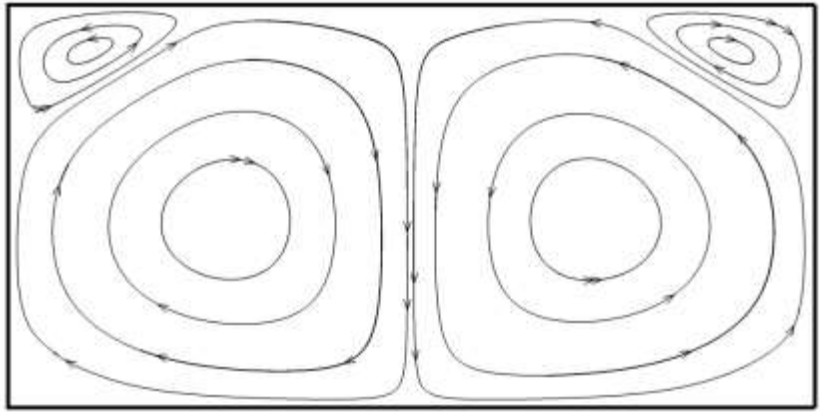

θ

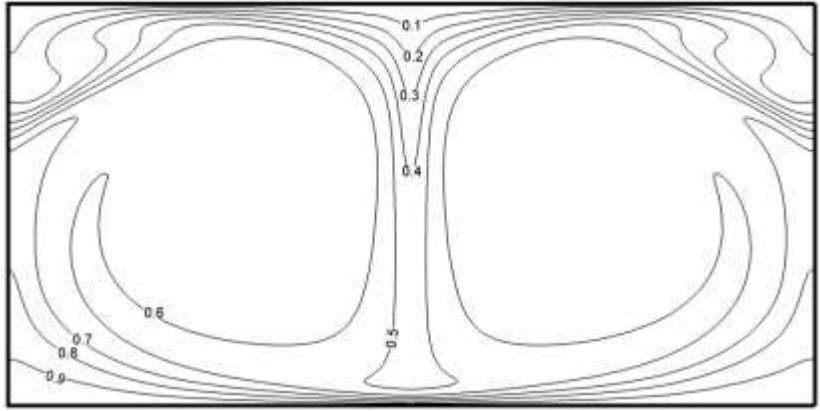

C

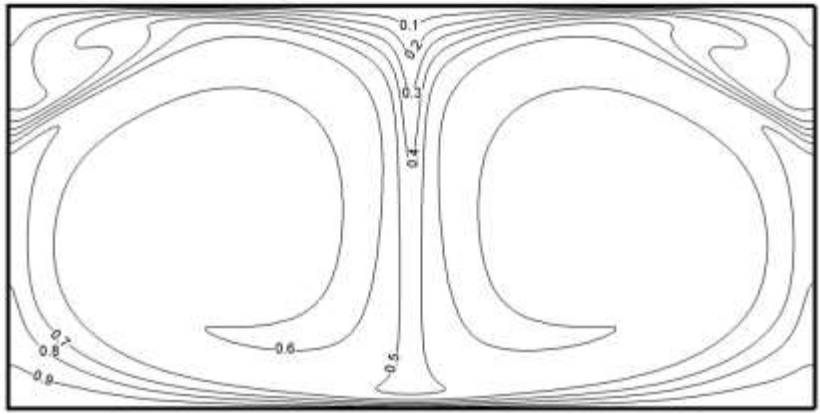

d) bicellular descending flow

**Figure 2** Steady-state solutions of flow flied, isotherms and isosolutes for double-diffusive convection ( $A = 0.5, Ra = 10^6, N_C = 0.5, S_r = D_f = 0.1$ )

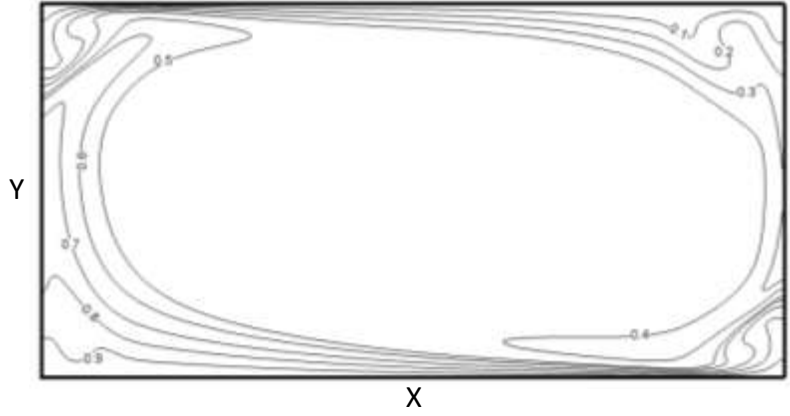

(a) $\tau = 350.548$

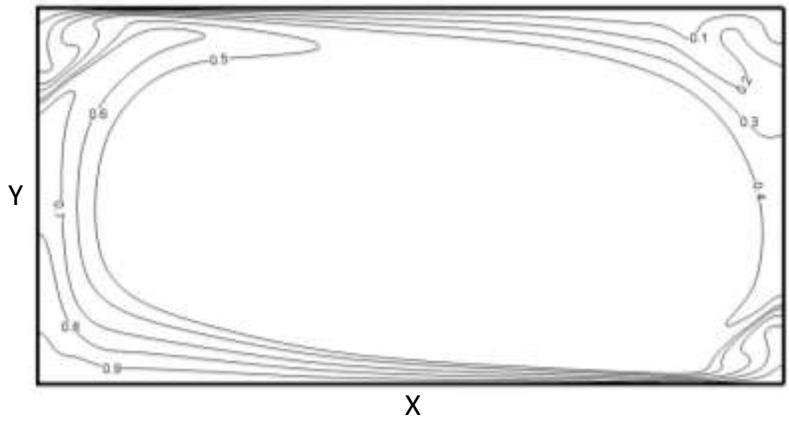

(b) $\tau = 350.608$

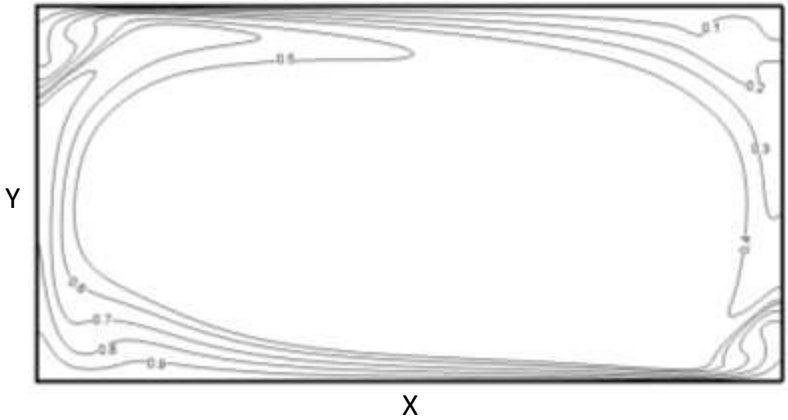

(c) $\tau = 350.688$

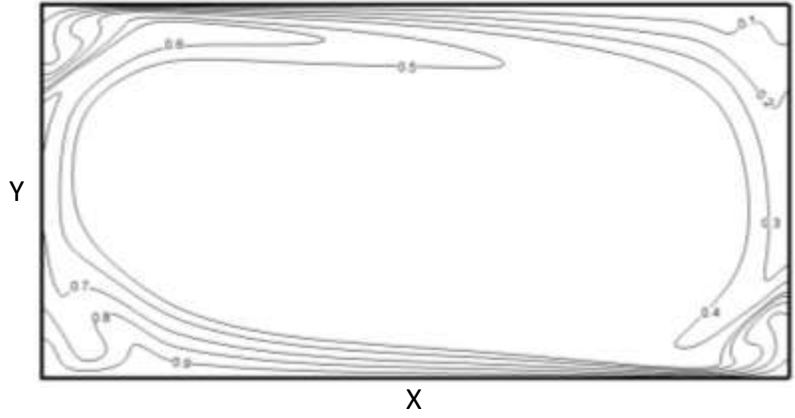

(d) $\tau = 350.768$

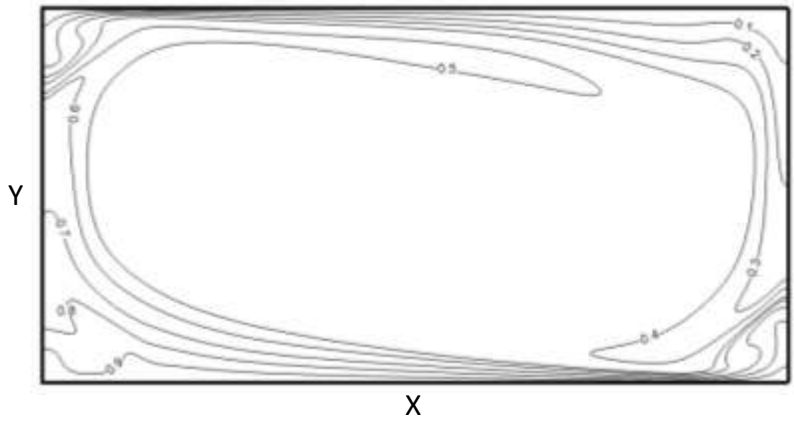

(e) $\tau = 350.870$

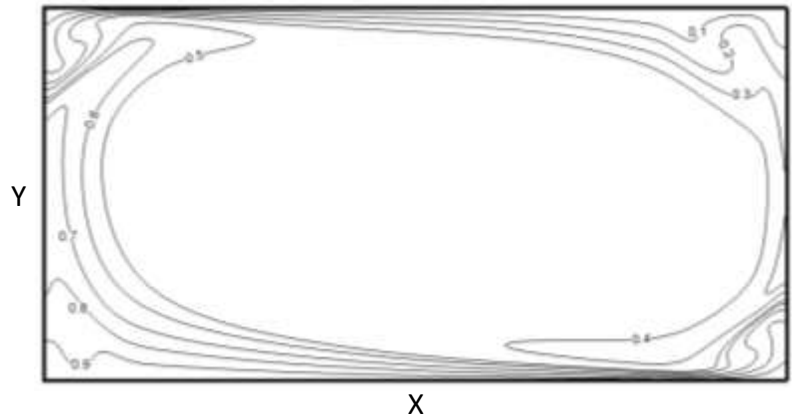

(f) $\tau = 350.922$

**Figure 3** Oscillatory solutions of temperature field for double-diffusive convection ( $A = 0.5, Ra = 10^6, N_C = 2.0, S_r = D_f = 0.1$ )

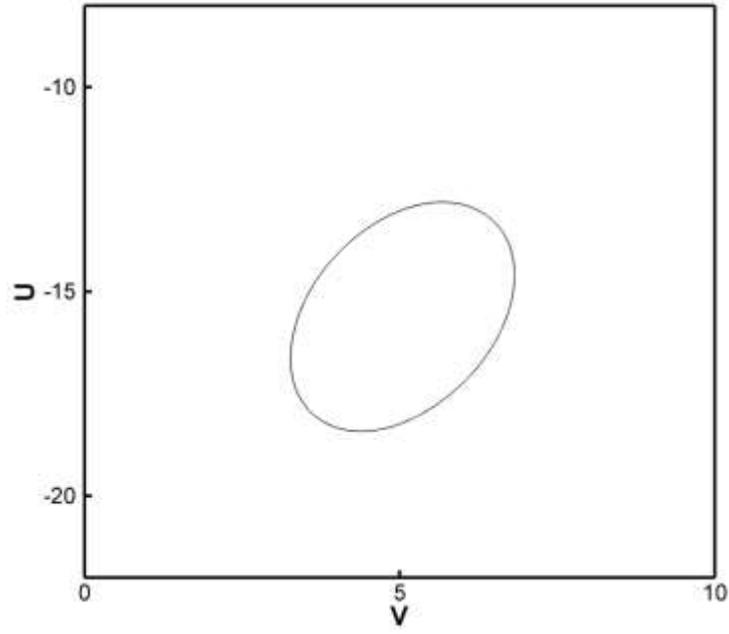

a) Phase-space trajectory of **U**

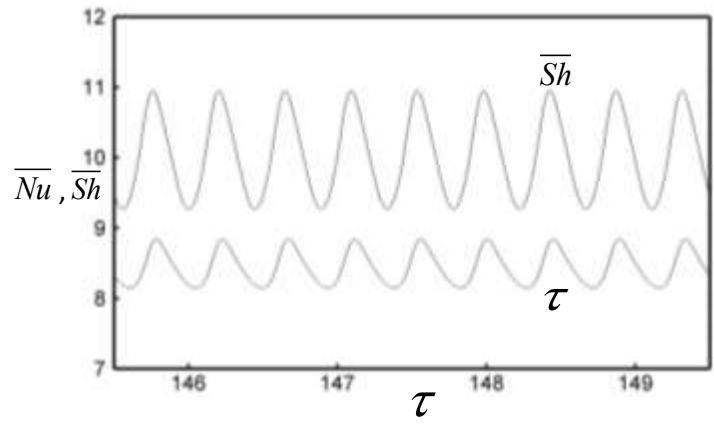

b) Time-evolution of heat and mass transfer

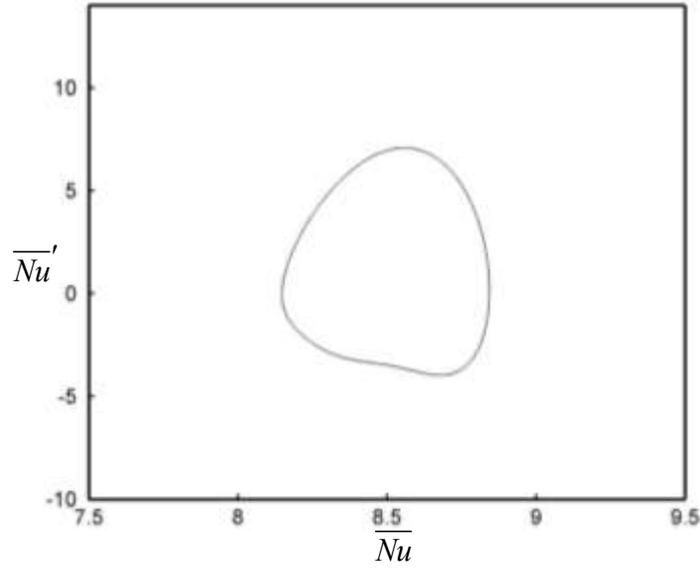

c) Phase-space trajectory of $\overline{Nu}$

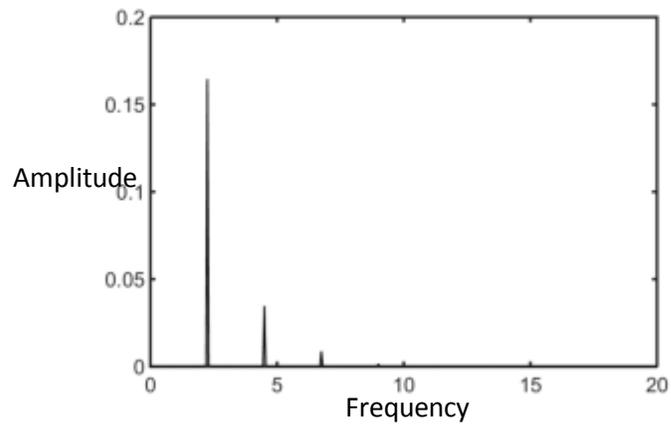

d) Fourier frequency spectrum of $\overline{Nu}$

**Figure 4** Oscillatory behaviors for $A=0.5$, $Ra=10^6$, $N_C=2.0$, $S_r=D_f=0.1$

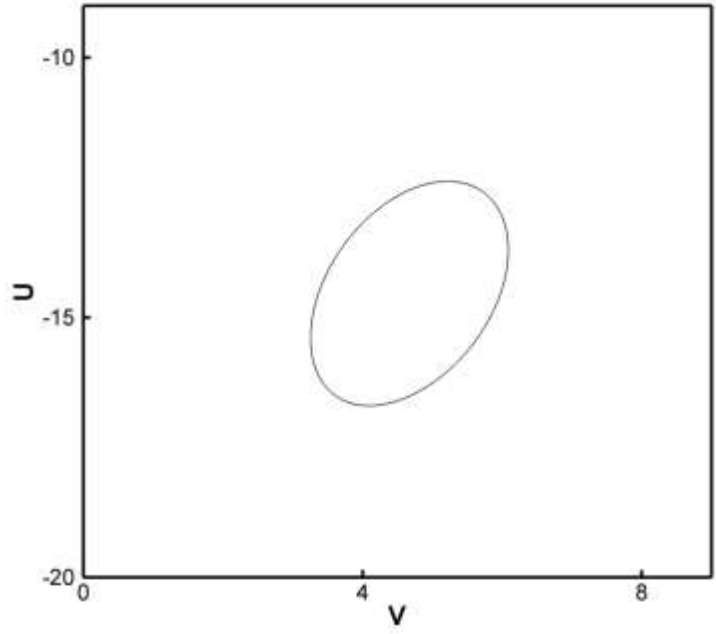

a) Phase-space trajectory of **U**

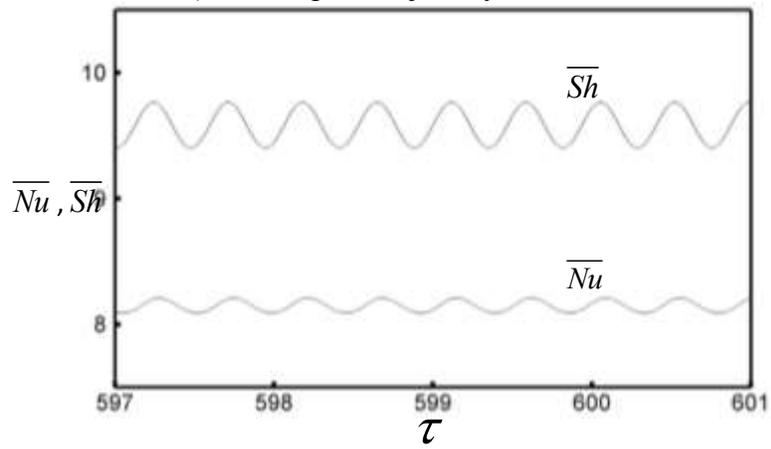

b) Time-evolution of heat and mass transfer

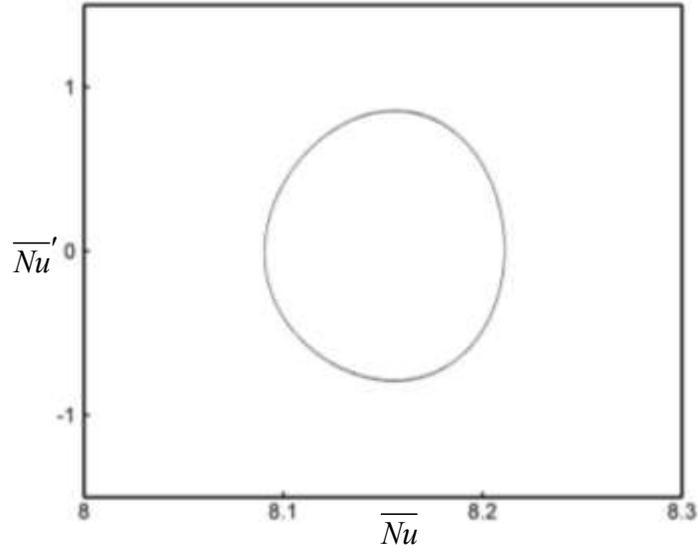

c) Phase-space trajectory of $\overline{Nu}$

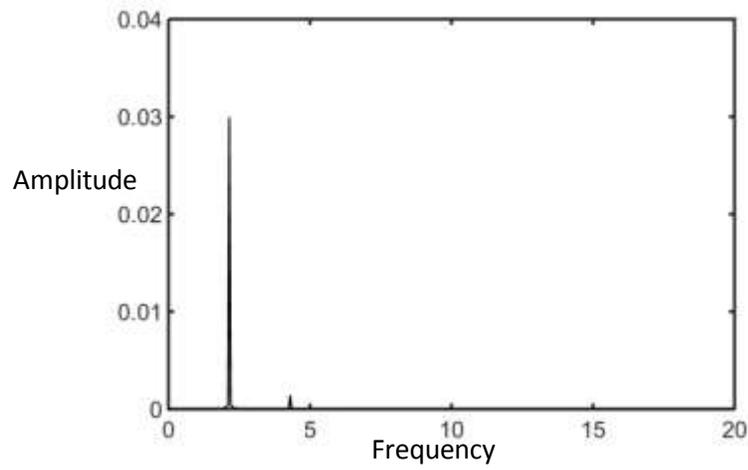

d) Fourier frequency spectrum of $\overline{Nu}$

**Figure 5** Oscillatory behaviors for $A = 0.5$, $Ra = 10^6$, $N_C = 1.5$, $S_r = D_f = 0.1$

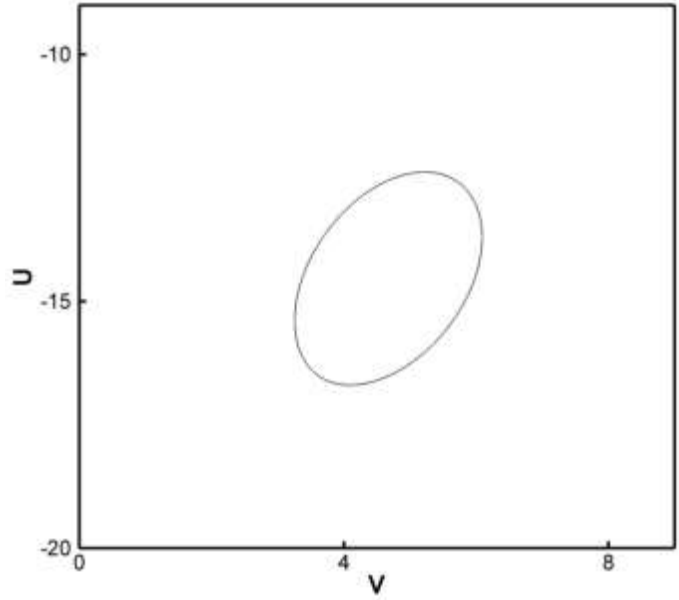
a) Phase-space trajectory of **U**

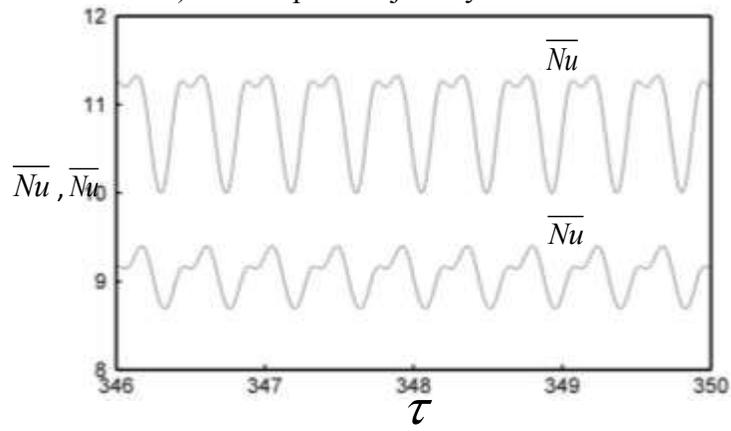
b) Time-evolution of heat and mass transfer

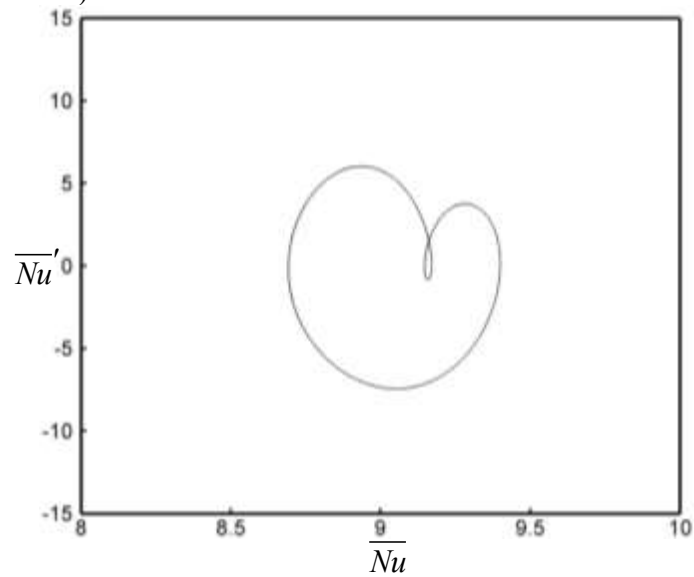

c) Phase-space trajectory of $\overline{Nu}$

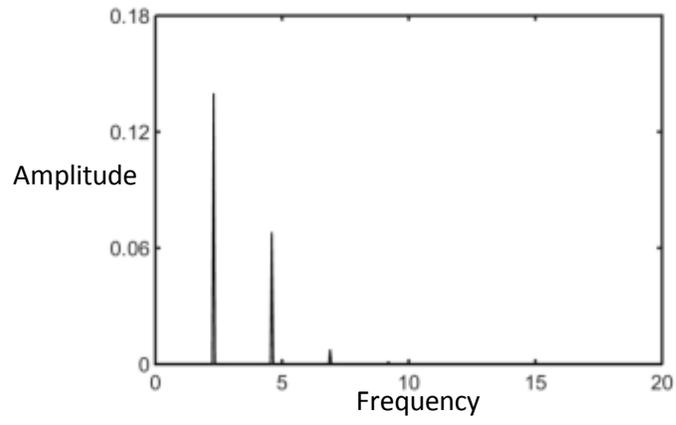

d) Fourier frequency spectrum of $\overline{Nu}$

**Figure 6** Oscillatory behaviors for $A=0.5, Ra=10^6, N_C=3.0, S_r=D_f=0.1$

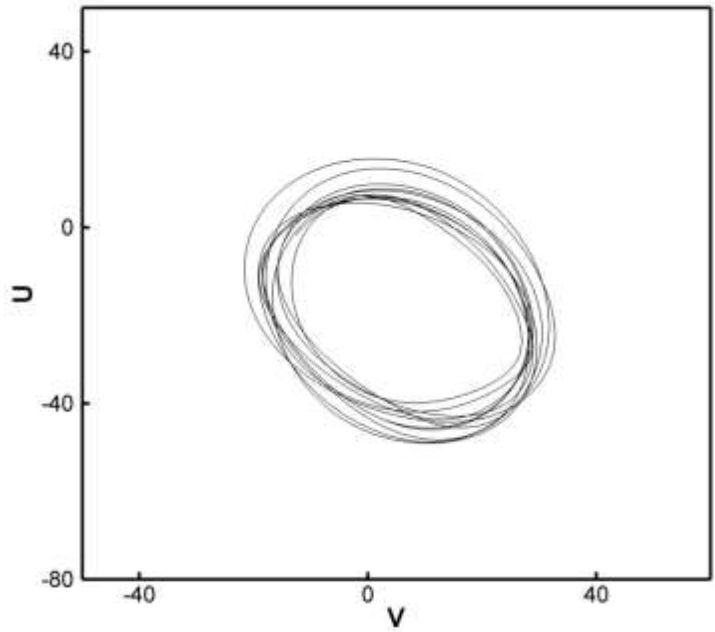

a) Phase-space trajectory of **U**

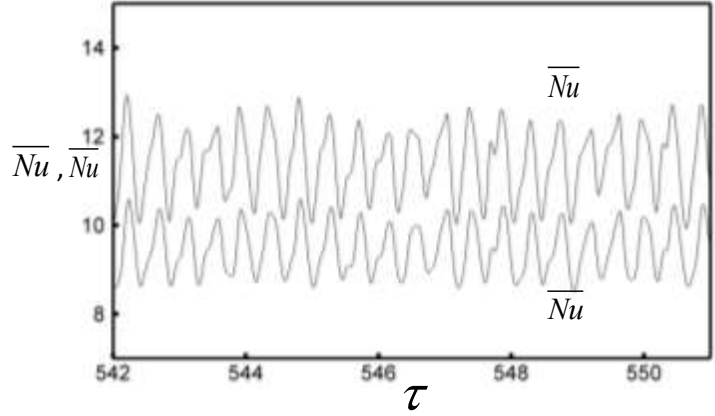

b) Time-evolution of heat and mass transfer

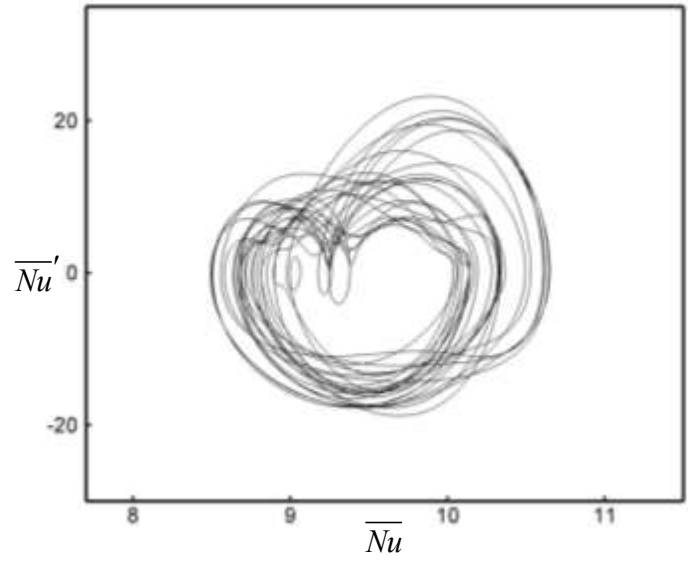

c) Phase-space trajectory of $\overline{Nu}$

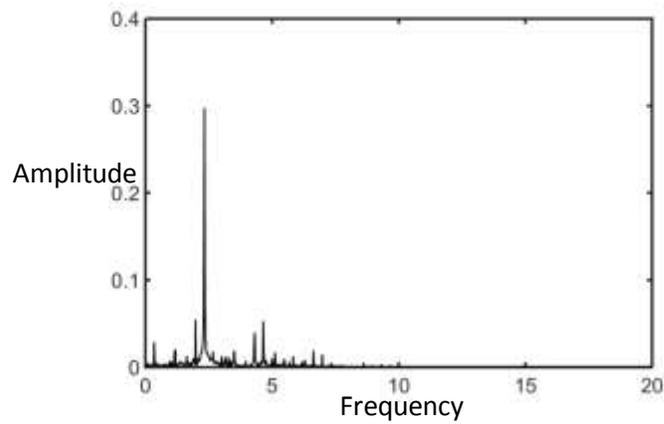

d) Fourier frequency spectrum of $\overline{Nu}$

**Figure 7** Oscillatory behaviors for $A = 0.5, Ra = 10^6, N_C = 4.0, S_r = D_f = 0.1$

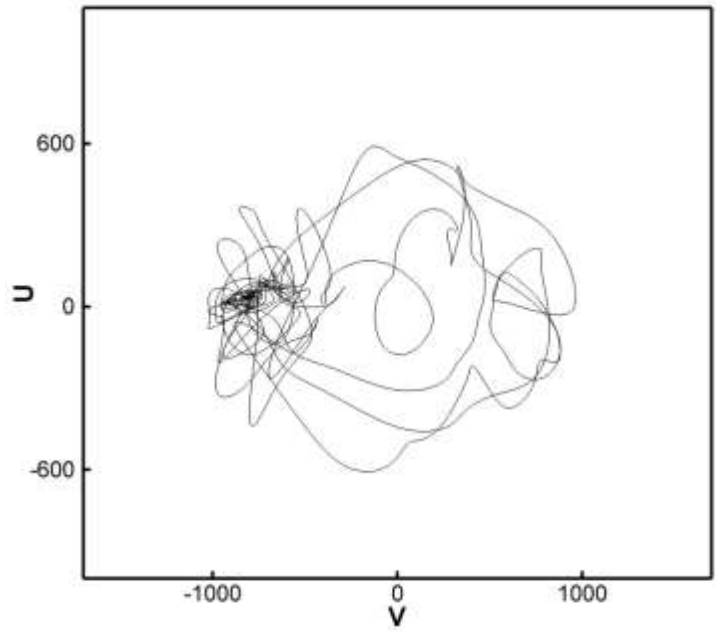

a) Phase-space trajectory of **U**

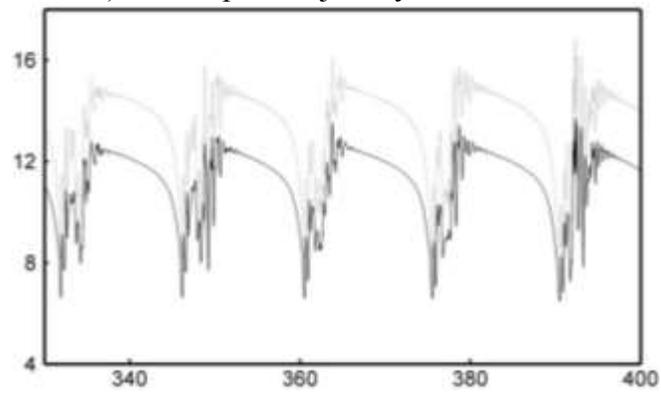

b) Time-evolution of heat and mass transfer

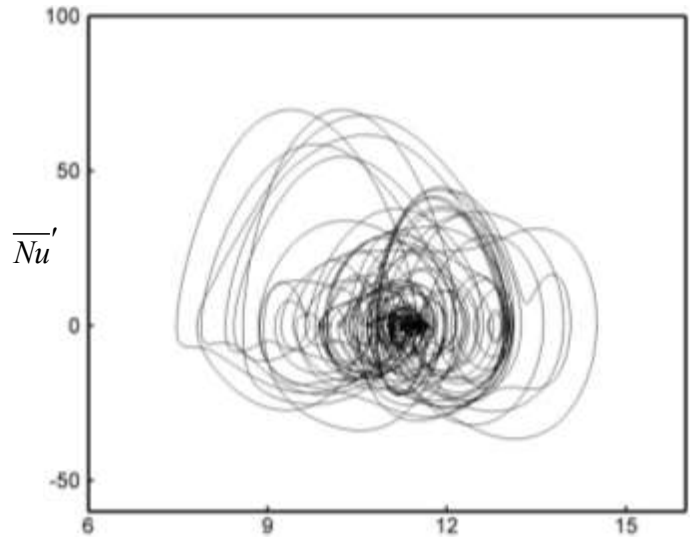

c) Phase-space trajectory of $\overline{Nu}$

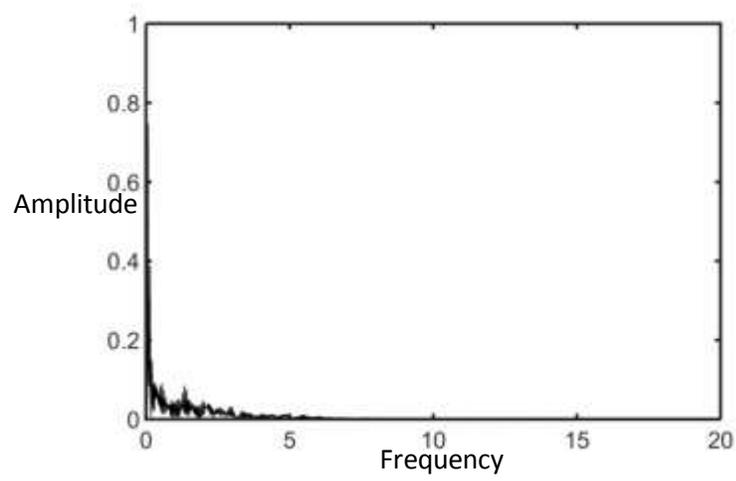

d) Fourier frequency spectrum of $\overline{Nu}$

**Figure 8** Oscillatory behaviors for $A = 0.5$, $Ra = 10^6$, $N_C = 5.0$, $S_r = D_f = 0.1$

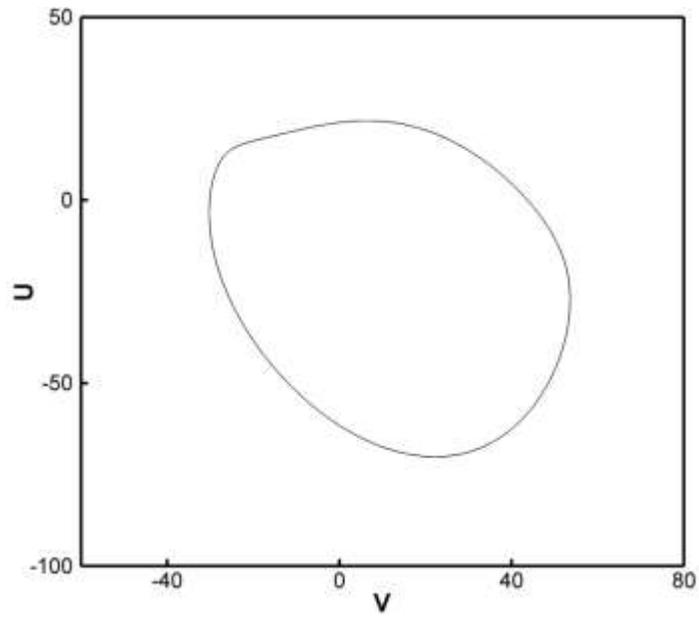
a) Phase-space trajectory of **U**

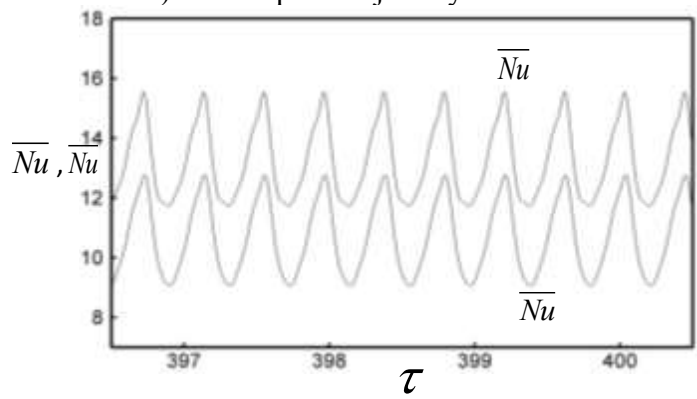
b) Time-evolution of heat and mass transfer

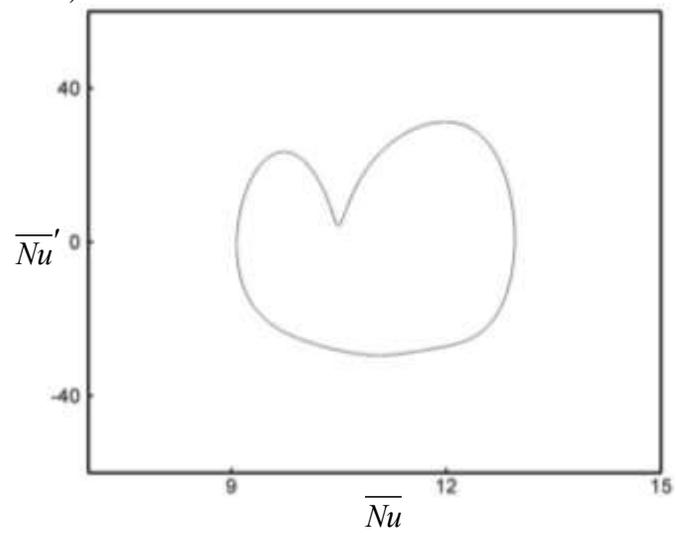
c) Phase-space trajectory of $\overline{Nu}$

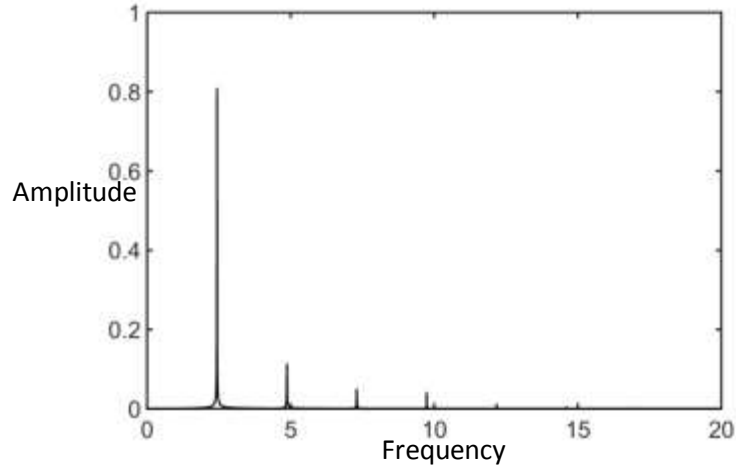

d) Fourier frequency spectrum of $\overline{Nu}$

**Figure 9** Oscillatory behaviors for $A=0.5, Ra=10^6, N_C=7.0, S_r=D_f=0.1$

Figures 5-9 show effects of buoyancy ratio on oscillatory double-diffusive convection in horizontal cavity. It can be seen that periodic self-sustained oscillation with fundamental frequency $FF=2.1498$ and double frequency $FF_2=4.2996$ in Fig. 5. And the fluctuation amplitude of heat transfer is 0.12, where the amplitude is defined as $\max(\overline{Nu})-\min(\overline{Nu})$, where $\max(\overline{Nu})$, $\min(\overline{Nu})$ are the maximum and the minimum values of $\overline{Nu}$, respectively. For $N_C=2.0$ and $N_C=3.0$, periodic oscillations with new FFs and larger fluctuation amplitude are shown in Figs. 4 and 6, where fundamental frequencies are 2.2498 and 2.2998, respectively. In other words, fundamental frequency and fluctuation amplitude increase with $N_C$ owing to stronger oscillation for larger buoyancy ratio. As $N_C$ increases to 4.0, the periodic oscillation evolves into quasi-periodic oscillation as shown in Fig. 7. For $N_C=5.0$, it is noted that fundamental frequency, sub-harmonics and background noises coexist as can be seen in Fig. 8; it demonstrates that the oscillatory convection changes into chaotic flow. However, the fluctuation amplitude still increases with buoyancy ratio, even if the oscillatory character of the double-diffusive convection changes. Further investigations for larger $N_C$ are carried out and it is unusual from Fig. 9 that the oscillation return to periodic oscillation with fundamental frequency $FF=2.4331$. Moreover, the fluctuation amplitude of $N_C=7.0$ is less than that of chaotic flow ($N_C=5.0$) due to change of the flow character. Therefore, the solution of double-diffusive convection evolves from steady-state convection-dominated, periodic oscillatory, quasi-periodic oscillatory into chaotic flow, and return to periodic oscillation as buoyancy ratio increases.

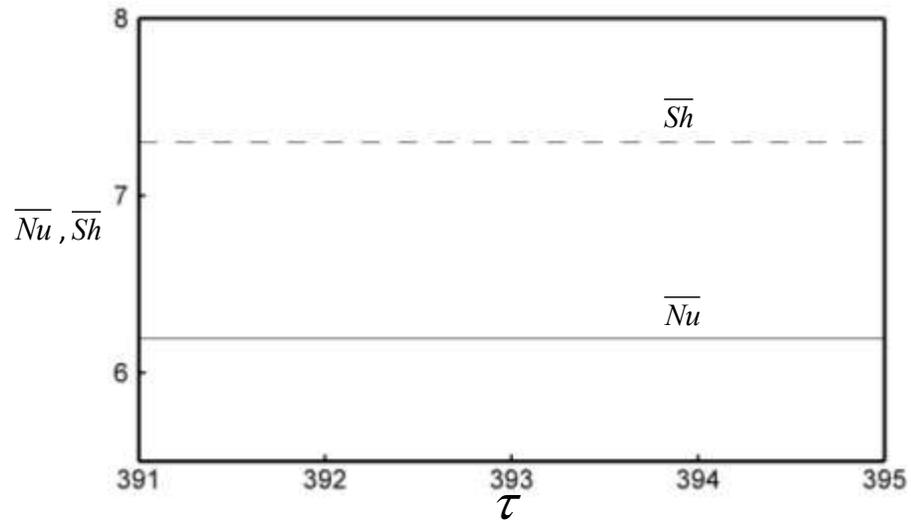

**Figure 10** Steady-state convection for $A = 0.5$, $Ra = 10^5$, $N_C = 3.0$, $S_r = D_f = 0.1$

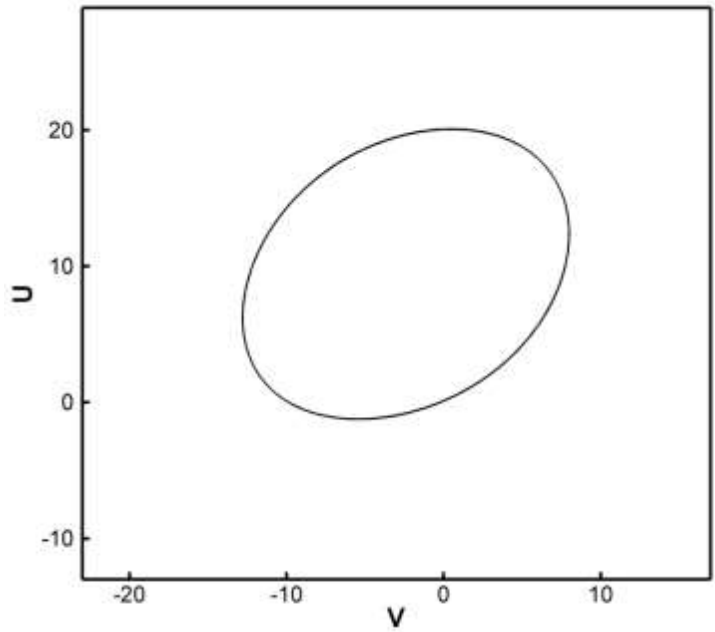

a) Phase-space trajectory of **U**

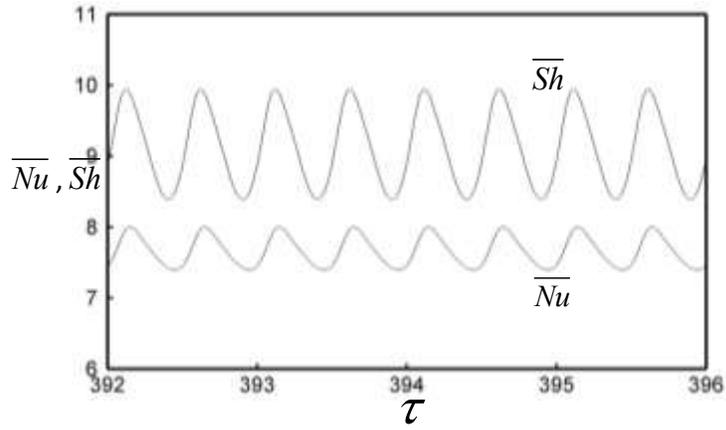

b) Time-evolution of heat and mass transfer

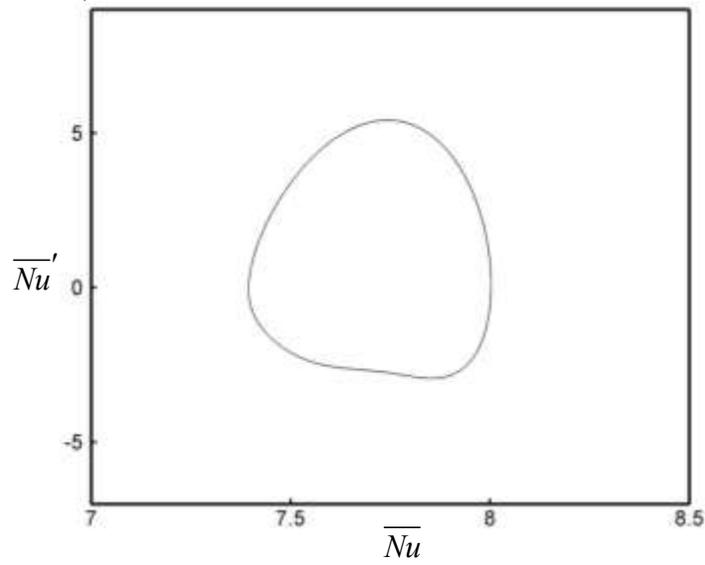

c) Phase-space trajectory of $\overline{Nu}$

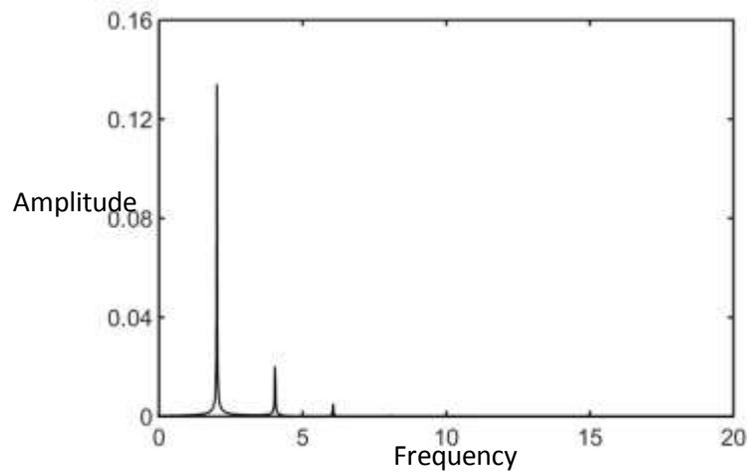

d) Fourier frequency spectrum of $\overline{Nu}$

**Figure 11** Oscillatory behaviors for $A = 0.5, Ra = 5\times10^5, N_C = 3.0, S_r = D_f = 0.1$

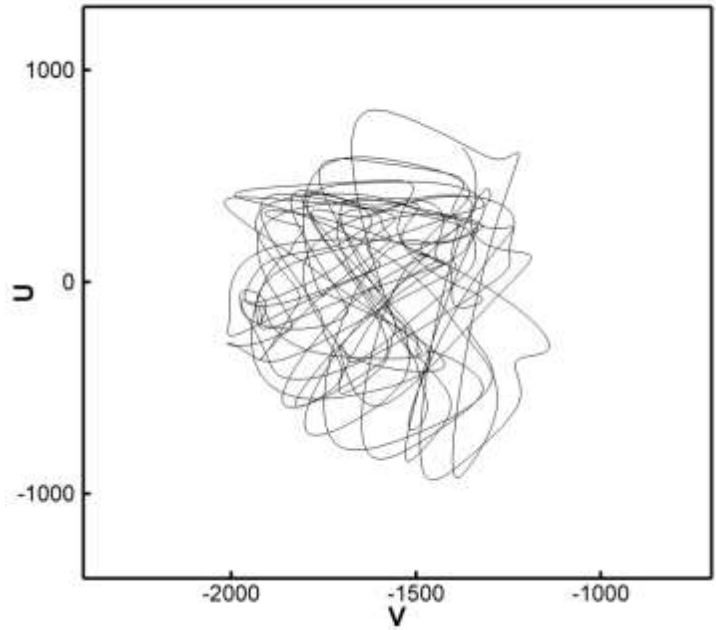

a) Phase-space trajectory of **U**

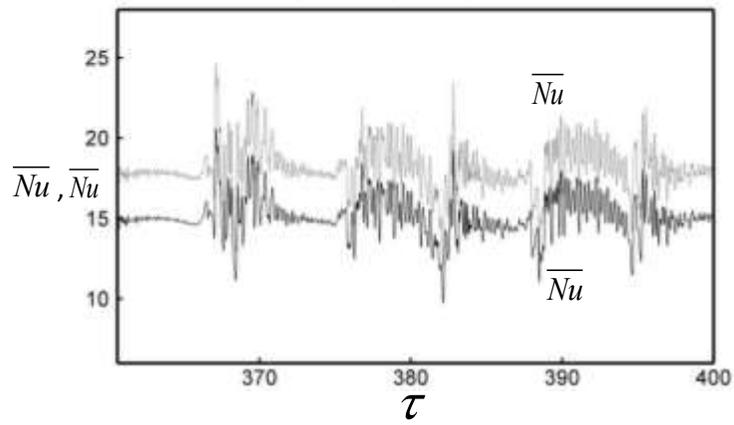

b) Time-evolution of heat and mass transfer

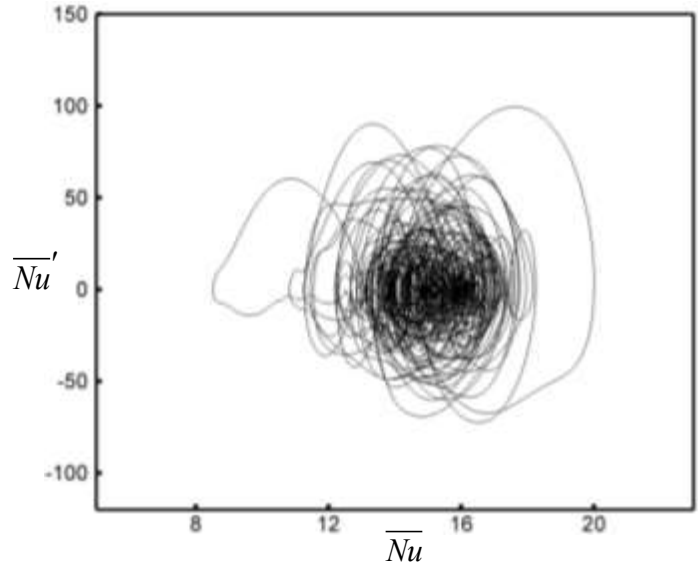

c) Phase-space trajectory of $\overline{Nu}$

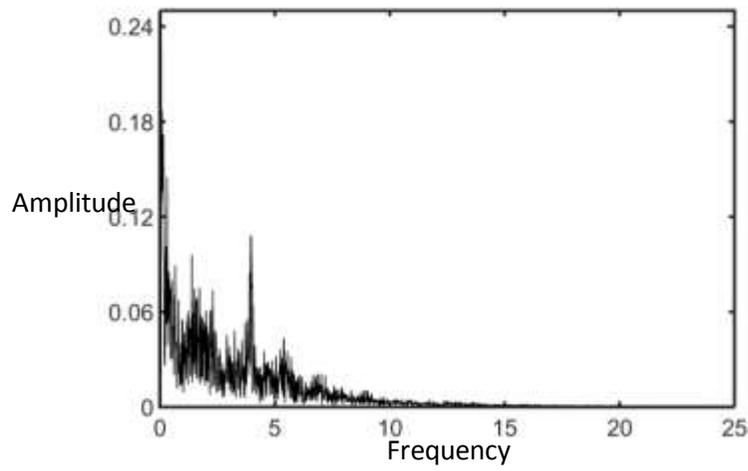

d) Fourier frequency spectrum of $\overline{Nu}$

**Figure 12** Oscillatory behaviors for $A=0.5, Ra=5\times10^6, N_C=3.0, S_r=D_f=0.1$

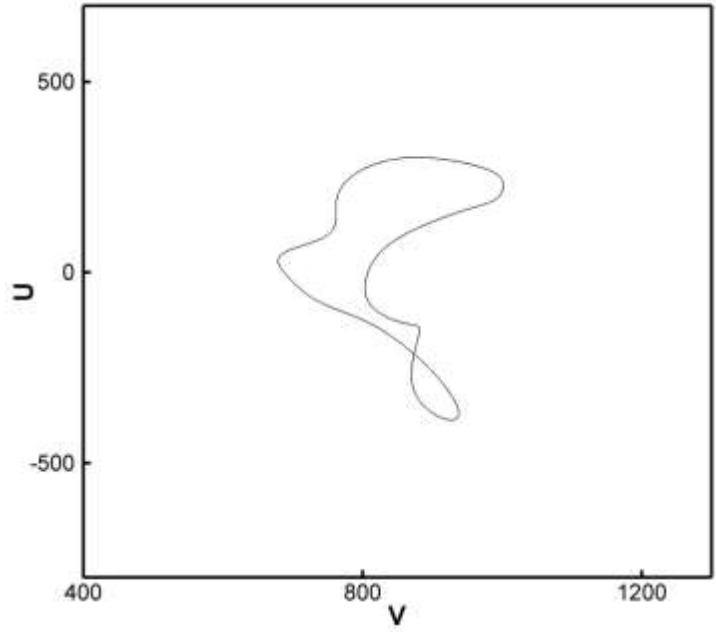

a) Phase-space trajectory of **U**

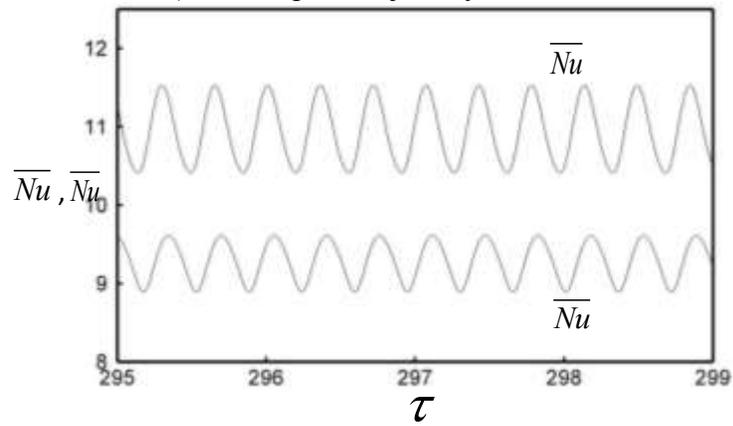

b) Time-evolution of heat and mass transfer

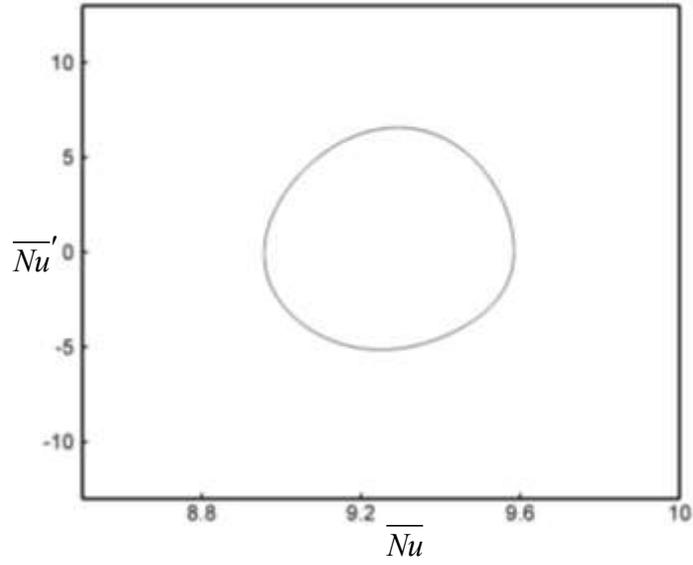

c) Phase-space trajectory of $\overline{Nu}$

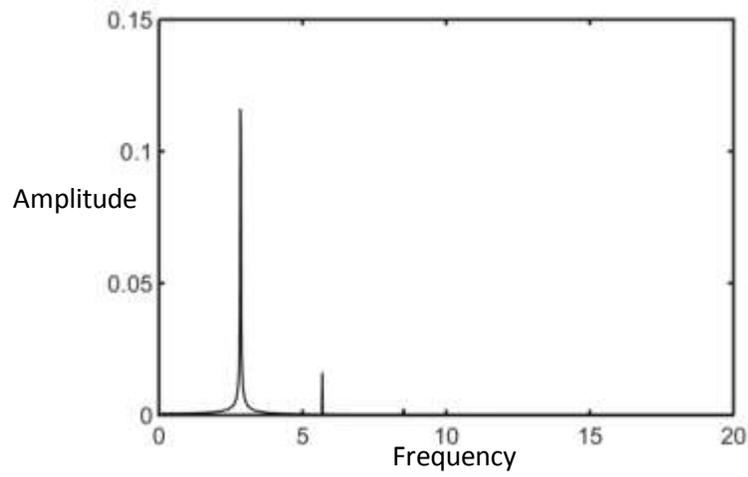

d) Fourier frequency spectrum of $\overline{Nu}$

**Figure 13** Oscillatory behaviors for $A = 0.25, Ra = 10^6, N_C = 3.0, S_r = D_f = 0.1$

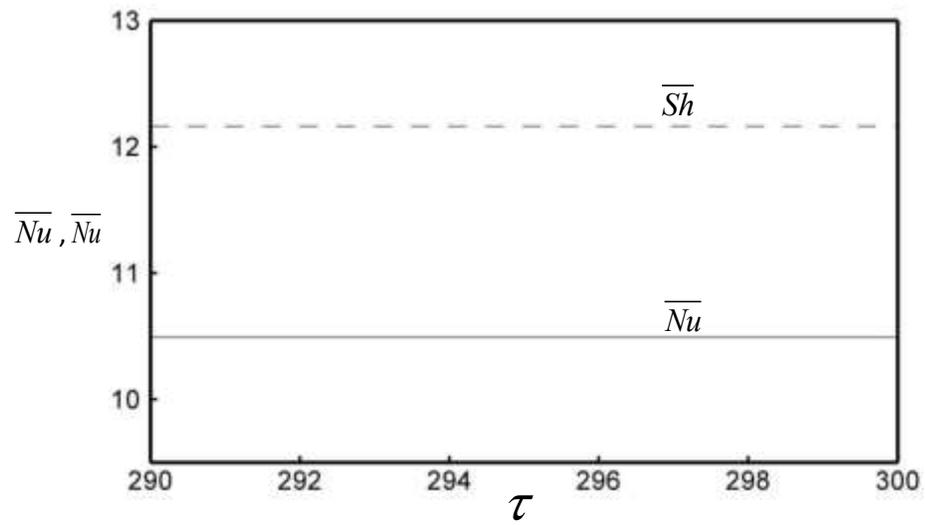

**Figure 14** Steady-state convection for $A = 0.125$, $Ra = 10^6$, $N_C = 3.0$, $S_r = D_f = 0.1$

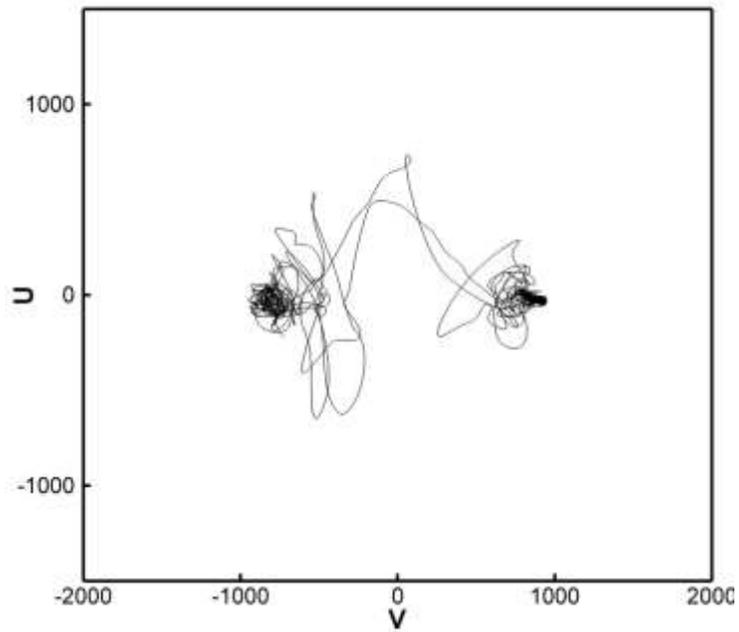

a) Phase-space trajectory of **U**

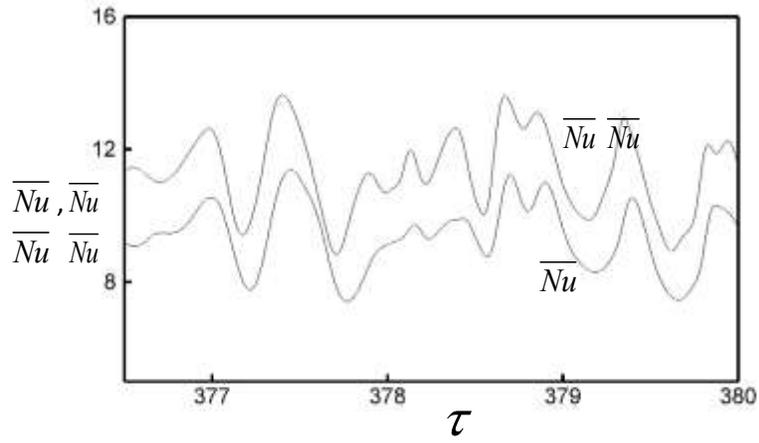

b) Time-evolution of heat and mass transfer

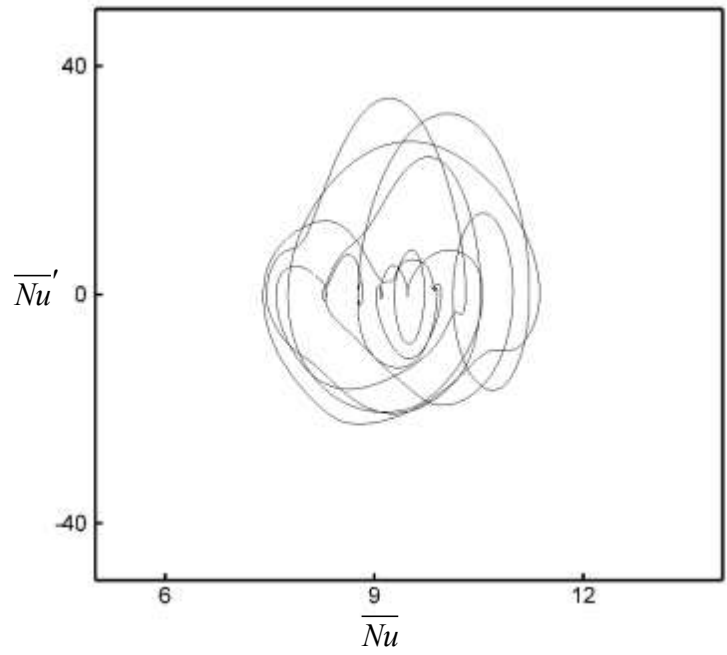

c) Phase-space trajectory of $\overline{Nu}$

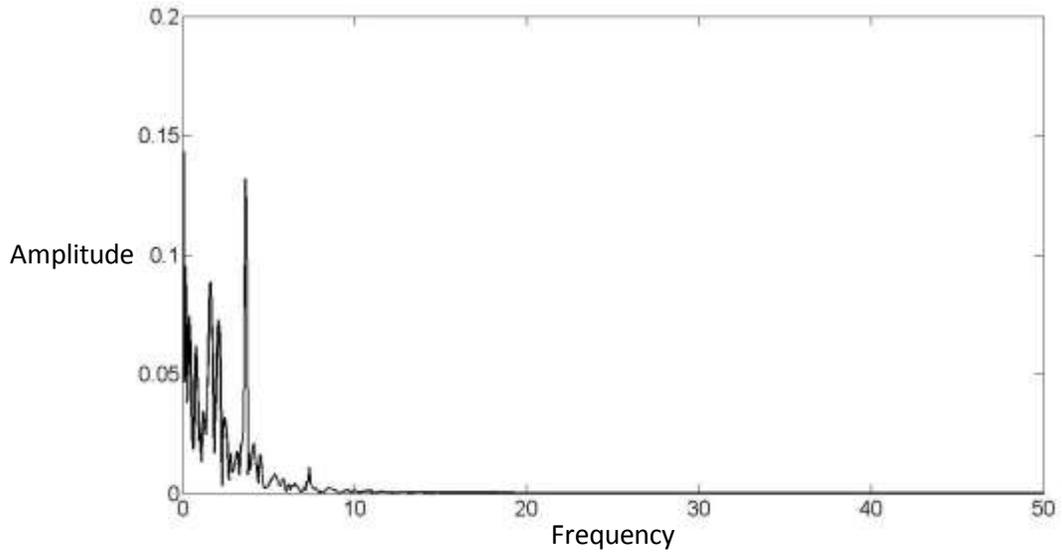

d) Fourier frequency spectrum of $\overline{Nu}$

**Figure 15** Oscillatory behaviors for $A=0.5, Ra=10^6, N_C=3.0, S_r=D_f=0.0$

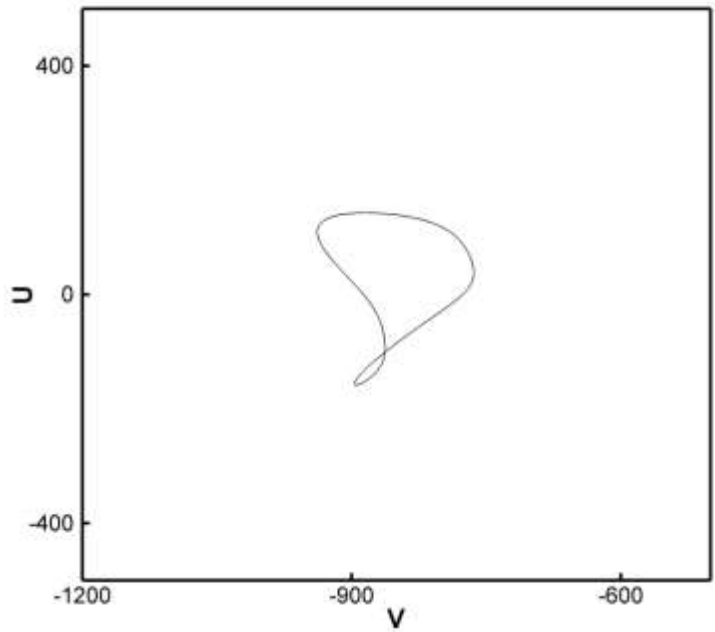

a) Phase-space trajectory of $\mathbf{U}$

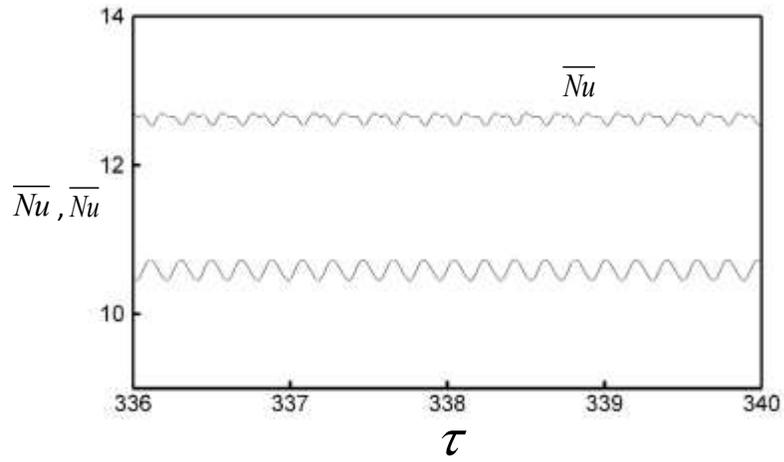

b) Time-evolution of heat and mass transfer

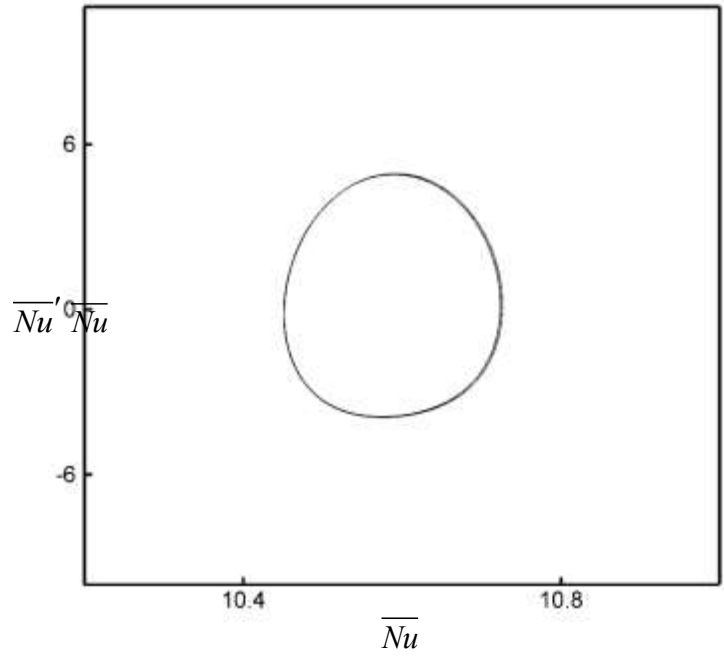

c) Phase-space trajectory of $\overline{Nu}$

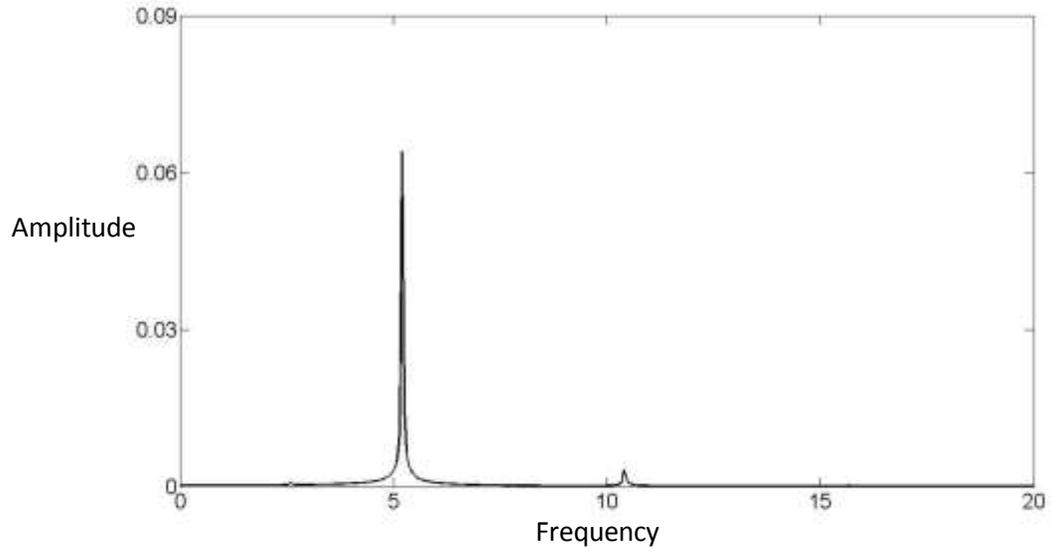

d) Fourier frequency spectrum of $\overline{Nu}$

**Figure 16** Oscillatory behaviors for $A=0.5, Ra=10^6, N_C=3.0, S_r=D_f=0.2$

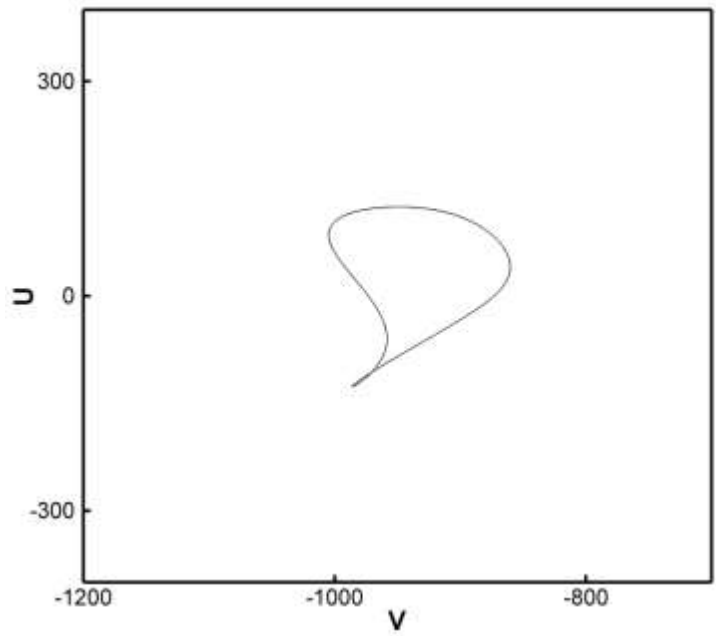

a) Phase-space trajectory of $U$

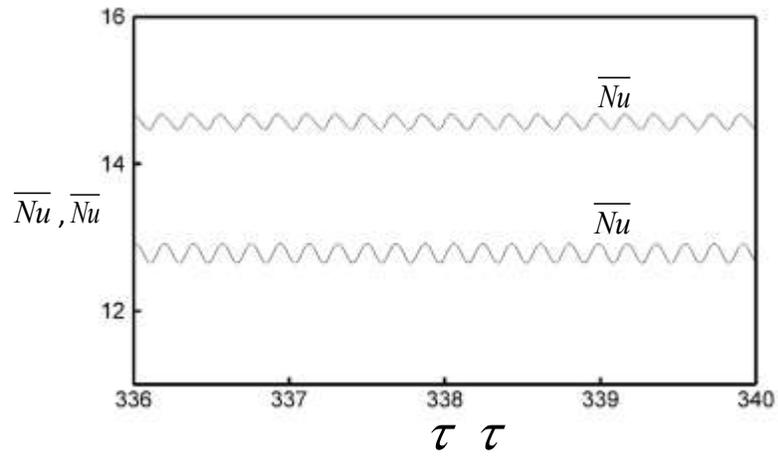

b) Time-evolution of heat and mass transfer

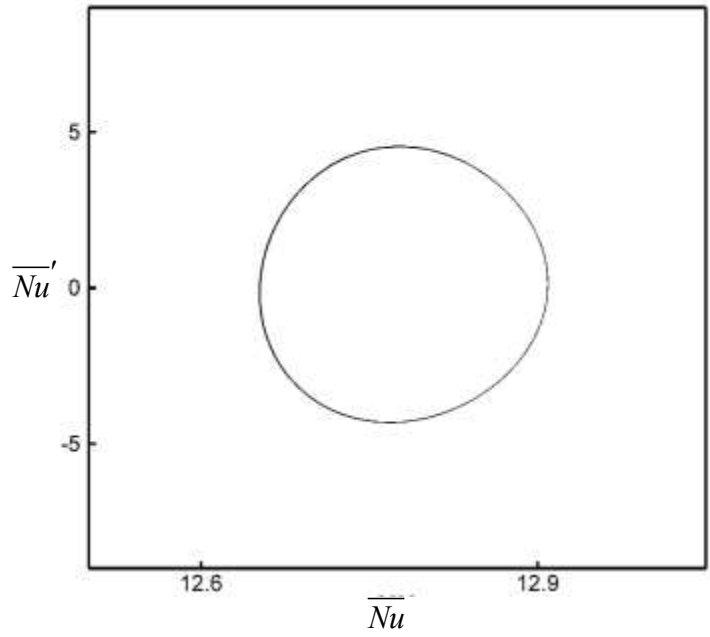

c) Phase-space trajectory of $\overline{Nu}$

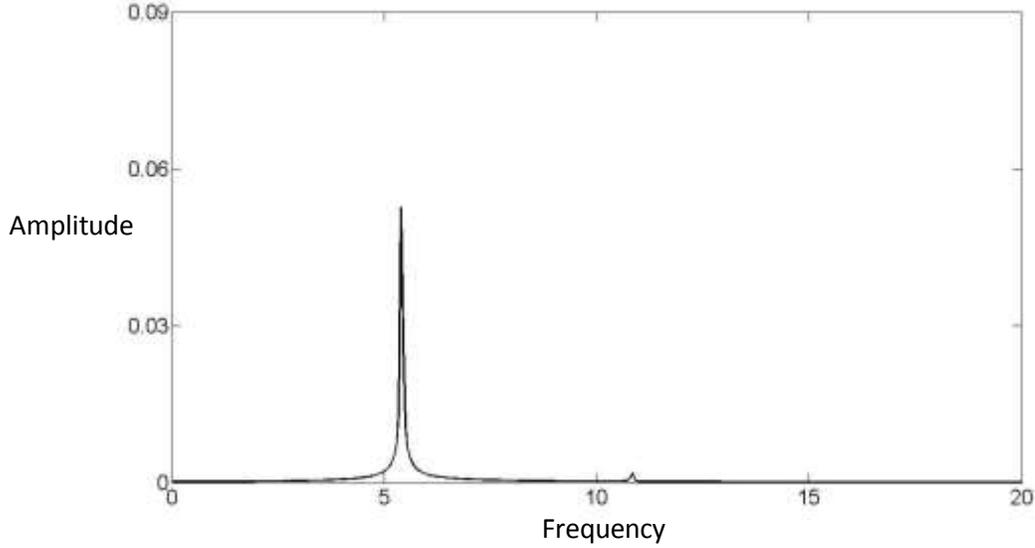

d) Fourier frequency spectrum of $\overline{Nu}$

**Figure 17** Oscillatory behaviors for $A = 0.5, Ra = 10^6, N_C = 3.0, S_r = D_f = 0.5$

Figures 10-12 show effect of Rayleigh number on the oscillatory double-diffusive convection in the horizontal cavity. It can be seen from Fig. 10 that $\overline{Nu}$ and $\overline{Sh}$ are almost unchanged with time because the system presents steady-state convection at $Ra = 10^5$. As $Ra$ increases, the steady-state convection changes into periodic oscillation as shown in Fig. 11, and finally develops into chaotic flow as shown in Fig. 12. The transition trendy of oscillatory convection based on increasing $Ra$ is similar to the effect of $N_C$. However, the return of periodic oscillation from chaos is not obtained as $Ra$ increases. In addition, more investigations of periodic oscillation demonstrate that fundamental frequency and fluctuation amplitude increase with $Ra$ and that the change trendy of $Ra$ is in line with that of $N_C$ because the increase of $Ra$ and $N_C$ are regarded as growth of the thermal and solutal buoyancies, respectively.

Figures 13-14 show the effect of aspect ratio on the oscillatory double-diffusive convection in horizontal cavity. It can be observed that the oscillatory convection develops from periodic (Figs. 6 and 13) into steady-state (Fig. 14) as $A$ decreases and that fluctuation amplitude decreases with $A$, finally it decreases into zero due to change of convection character into steady state. Meanwhile, it can be seen from Figs. 6(b), 13(b) and 14 that $\max(\overline{Nu})$ and $\max(\overline{Sh})$ (which are the maximums of $\overline{Nu}$ and $\overline{Sh}$ correspondence to dimensionless time) increase as the oscillation return to steady-state convection; more studies of one period show that heat and mass transfer enhances with decreasing $A$ ($A \in [0.125, 0.5]$). More investigations for different aspect ratios show that fundamental frequency increases at first and then decreases as $A$ decreases due to changes of flow structure and character.

Figures 15-17 show the Soret and Dufour effects on the oscillatory double-diffusive convection in horizontal cavity. It can be seen that the convection is chaotic without the Soret and Dufour effects for $A = 0.5, Ra = 10^6, N_C = 3.0$ in Fig. 15. As Soret and Dufour effects increase, the oscillatory convection evolves from chaotic into periodic as shown in Figs. 6, 16 and 17. The fundamental

frequency, $\max(\overline{Nu})$ and $\max(\overline{Sh})$ increase because the increasing Soret effect enhances heat transfer and the increasing Dufour effect enhances mass transfer enhance. And the increasing heat and mass transfer enhance each other so that it make the oscillatory periodicity decrease and fundamental frequency increase.

## 5. Conclusions

Systematic investigations of oscillatory double-diffusive convection with Soret and Dufour effects in horizontal cavity for different Rayleigh numbers, buoyancy ratios and aspect ratios have been carried out. The numerical results show that double-diffusive convection develops from steady-state convection-dominated, periodic oscillatory, quasi-periodic oscillatory to chaotic flow as buoyancy ratio or Rayleigh number increases. Moreover, system return to periodic oscillation from chaos as buoyancy ratio continues to increase. Fundamental frequency and fluctuation amplitude increase with Rayleigh number or buoyancy ratio. As aspect ratio decreases, the oscillatory convection evolves from periodic into steady-state. Meanwhile, fundamental frequency increases at first and then decreases while fluctuation amplitude decreases with aspect ratio. As Soret and Dufour effects increase, the oscillatory convection changes from chaotic into periodic. Meanwhile, fundamental frequency, $\max(\overline{Nu})$ and $\max(\overline{Sh})$ increase with Soret and Dufour effects because the increasing Soret and Dufour effects make heat transfer and mass transfer enhance.

**Acknowledgment**

This work is supported by Chinese National Natural Science Foundations under Grants 51129602 and 51476103, Innovation Program of Shanghai Municipal Education Commission 14ZZ134, Innovation Fund Project for Graduate Student of Shanghai JWCXSL1301, and SQI Commonweal Project NO.2012-12.

**Nomenclature**

| | |
|---|---|
| $c$ | concentration, $kg/m^3$ |
| $C$ | dimensionless concentration |
| $D$ | diffusion coefficient, $m^2/s$ |
| $D_f$ | Dufour coefficient |
| $g$ | gravitational acceleration, $m^2/s$ |
| $Le$ | Lewis number |
| $N_C$ | buoyancy ratio |
| $\overline{Nu}$ | average Nusselt number |
| $\overline{Nu}'$ | Derivative of average Nusselt number with respect to $\tau$ |
| $P$ | dimensionless pressure |
| $\Pr$ | Prandtl number |
| $Ra$ | Rayleigh number |
| $\overline{Sh}$ | average Sherwood number |
| $S_r$ | Soret coefficient |
| $\theta$ | dimensionless temperature |

**Greek symbols**

| | |
|---|---|
| $\tau$ | dimensionless time |
| $\kappa_{TC}$ | Duffour coefficient, $m^5 \cdot K \cdot kg^{-1} \cdot s^{-1}$ |
| $\kappa_{CT}$ | Soret coefficient, $kg \cdot m^{-1} \cdot K^{-1} \cdot s^{-1}$ |
| $\beta_T$ | thermal volumetric expansion, $K^{-1}$ |
| $\beta_c$ | solute volumetric expansion, $m^3/Kg$ |
| $\nu$ | kinematical viscosity, $m^2/s$ |

**Subscripts**

| | |
|---|---|
| 0 | initial condition |
| h | high temperature or concentration condition |
| l | low temperature or concentration |